\documentclass[prd,twocolumn,floatfix,amsmath,nofootinbib,amssymb,floatfix]{revtex4}
\usepackage{graphicx,color,dcolumn,booktabs,bm}
\usepackage{longtable,lscape}
\usepackage{pdfpages}
\usepackage{txfonts}
\usepackage{overpic}
\usepackage{amssymb}
\usepackage{makecell}
\usepackage{indentfirst}
\usepackage{feynmf}   
\usepackage{slashed}  
\usepackage{cases}
\usepackage{color}
\usepackage{multirow}
\usepackage{threeparttable}
\usepackage{epstopdf}
\usepackage{enumerate}
\usepackage{subfigure}
\usepackage{diagbox}
\usepackage{graphicx,color,dcolumn,booktabs,bm}
\usepackage{mathrsfs}
\usepackage{cancel}
\usepackage{float}
\usepackage[misc]{ifsym}
\usepackage[colorlinks,
            citecolor=blue,
            anchorcolor=red,
            menucolor=red,
            linkcolor=red,
            filecolor=red,
            runcolor=red,
            urlcolor=blue,
            frenchlinks=red]{hyperref}
\usepackage{array}
\usepackage{booktabs}

\begin{document}
\title{Analysis of the semileptonic decays $\Sigma_b\to\Sigma_cl\bar{\nu}_l$, $\Xi'_b\to\Xi'_cl\bar{\nu}_l$ and $\Omega_b\to\Omega_cl\bar{\nu}_l$ in QCD sum rules}
\author{Jie Lu$^{1}$}
\email{l17693567997@163.com}
\author{Guo-Liang Yu$^{2}$}
\email{yuguoliang2011@163.com}
\author{Dian-Yong Chen$^{1,3}$}
\email{chendy@seu.edu.cn}
\author{Zhi-Gang Wang$^{2}$}
\email{zgwang@aliyun.com}
\author{Bin Wu$^{1}$}

\affiliation{$^1$ School of Physics, Southeast University, Nanjing 210094, People's Republic of
China\\$^2$ Department of Mathematics and Physics, North China
Electric Power University, Baoding 071003, People's Republic of
China\\$^3$ Lanzhou Center for Theoretical Physics, Lanzhou University, Lanzhou 730000, People's Republic of
China}
\date{\today }

\begin{abstract}
In this article, the electroweak transition form factors of $\Sigma_b\to\Sigma_c$, $\Xi'_b\to\Xi'_c$ and $\Omega_b\to\Omega_c$ are analyzed within the framework of three-point QCD sum rules. In phenomenological side, all possible couplings of interpolating current to hadronic states are considered, and the Dirac structure dependence on the form factors is systematically eliminated. In QCD side, our calculation incorporates both the perturbative part and the contributions from vacuum condensates up to dimension 8. This systematic inclusion of higher-dimensional terms accounts for a broader set of Feynman diagrams, thereby enhancing the comprehensiveness and reliability of the operator product expansion. Using the obtained form factors, we study the partial widths of semileptonic decays $\Sigma_b\to\Sigma_cl\bar{\nu}_l$, $\Xi'_b\to\Xi'_cl\bar{\nu}_l$ and $\Omega_b\to\Omega_cl\bar{\nu}_l$ ($l=e$, $\mu$ and $\tau$). The results indicate that these decay widths approximately satisfy SU(3) flavor symmetry. Next, we calculate the branching ratios for the decay process $\Omega_b\to\Omega_cl\bar{\nu}_l$ and compare them with the results from other collaborations. Furthermore, the lepton universality ratios and some asymmetry parameters of these decay processes are also analyzed, which provide information for the study of new physics. We hope that these results will serve as a useful reference for future theoretical and experimental studies of weak decays involving heavy flavor baryons.
\end{abstract}


\maketitle

\section{Introduction}\label{sec1}
The investigation of the weak decay processes of heavy flavor hadrons is of great significance in both theory and experiment as it is a suitable method to test the standard model (SM) and find the new physics beyond the SM. In particular, the semileptonic decays provide a cleaner theoretical framework for studying weak transition dynamics, owing to the absence of final-state strong interactions. Compared with meson decays, the semileptonic decays of heavy flavor baryons offer additional insights into the dynamics of heavy quarks within a three body system, albeit with increased theoretical complexity. The single heavy baryons contain one heavy (charmed or bottom) quark and two light quarks, which is an ideal testing ground for the heavy quark effective theory (HQET). In the quark model, these single heavy baryons are classified by their flavor wave functions. With the heavy quark wave function fixed, the remaining two light quarks form a tensor product representation $3_F\otimes3_F$, which decomposes into the flavor anti-triplet $\bar{3}_F$ and the flavor sextet $6_F$. The anti-triplet includes $\Lambda^+_c(\Lambda^0_b)$ and $\Xi^{+(0)}_c(\Xi^{0(-)}_b)$, while the sextet comprises $\Sigma^{++(+,0)}_c(\Sigma^{+(0,-)}_b)$, $\Xi'^{+(0)}_c(\Xi'^{0(-)}_b)$ and $\Omega^{0}_c(\Omega^{-}_b)$.

Experimentally, numerous single heavy baryons have been identified in recent years by collaborations such as LHCb, Belle, BABAR and CLEO~\cite{ParticleDataGroup:2024cfk}. To date, more than 60 decay channels involving both single charmed and bottom baryons have been observed~\cite{ParticleDataGroup:2024cfk}. These discoveries have spurred significant theoretical interest in understanding their internal structure and decay properties. Theoretically, key properties including the mass spectra, Regge trajectories, magnetic momentum and strong (weak) decays of single heavy baryons have been extensively studied using a variety of approaches, such as QCD sum rules (QCDSR)~\cite{Wang:2009cr,Wang:2010it,Neishabouri:2025abl,Neishabouri:2024gbc,Luo:2025sns,Lu:2025gol,Yu:2026tbk}, Light-cone QCD sum rules (LCSR)~\cite{Wang:2008ni,Wang:2009hra,Khodjamirian:2011jp,Wang:2015ndk,Aliev:2016xvq,Aliev:2022gxi,Shi:2024plf,Luo:2025pzb,Aliev:2025cko}, various quark models~\cite{Cheng:1995fe,Ivanov:1999pz,Zhao:2018zcb,Yu:2022ymb,Li:2022xtj,Li:2024zze,Zhang:2025pde,Patel:2025rhg} and others~\cite{Xu:1992hj,Du:2011nj,Han:2020sag,Ivanov:1998ya,Sheng:2020drc,Yu:2023bxn,Wang:2022zcy,Zhou:2023xxs,Li:2025rsm,Neishabouri:2026mjn,Amiri:2025zdj}. Among these studies, semileptonic decays are important as a sensitive probe of QCD in the non-perturbative region. In particular, decays governed by the $b \to c$ transition allow for stringent tests of HQET and SU(3) flavor symmetry and provide an independent extraction of the Cabibbo-Kobayashi-Maskawa (CKM) matrix element $V_{cb}$.  

The weak decays of ground state flavor anti-triplet single bottom baryons $\Lambda_b$ and $\Xi_b$ have been extensively studied both experimentally and theoretically. For instance, the semileptonic decay channel $\Lambda_b\to\Lambda_c l\bar{\nu}_l$ has been measured by LHCb collaboration~\cite{LHCb:2017vhq,LHCb:2022piu} and studied in various theoretical frameworks~\cite{Gutsche:2014zna,Zhao:2020mod,Zhang:2022bvl}. Similarly, the decay $\Xi_b\to\Xi_c l\bar{\nu}_l$ has been also studied by various theoretical approaches~\cite{Zhao:2020mod,Neishabouri:2025abl,Faustov:2018ahb}. In contrast, the investigations of weak decays for flavor sextet single bottom baryons remain relatively unexplored. Due to the large branching ratios of strong decay channels such as $\Sigma^-_b\to\Lambda^0_b\pi^-$ and $\Xi'^-_b\to\Xi^0_b\pi^-$, the ground state $\Sigma^-_b$ and $\Xi'^-_b$ baryons have relatively large decay widths. The total decay widths of $\Sigma^-_b$ and $\Xi'^-_b$ are given by the Particle Data Group (PDG), which are $5.3\pm0.5$ and $0.03\pm0.032$ MeV, respectively~\cite{ParticleDataGroup:2024cfk}. Consequently, their semileptonic branching ratios are too small to be observed with current experimental sensitivity. However, the situation differs from the ground state $\Omega^-_b$ baryon. Its lifetime is given as $(1.64\pm0.16)\times10^{-12}$ s~\cite{ParticleDataGroup:2024cfk} which is close to that of $\Lambda^0_b$ and that of $\Xi^-_b$. Since the semileptonic decay channels are important for $\Lambda_b$ and $\Xi_b$ baryons, the study of $\Omega_b$ semileptonic decays is also particularly valuable. Moreover, the ground state flavor sextet single heavy baryons share similar internal structures, a systematic study of their properties offers a meaningful opportunity to probe SU(3) flavor symmetry and its breaking patterns in the single heavy baryon sector. 

The semileptonic decays $\Sigma_b\to\Sigma_c l\bar{\nu}_l$, $\Xi'_b\to\Xi'_c l\bar{\nu}_l$ and $\Omega_b\to\Omega_c l\bar{\nu}_l$ are all dominated by the quark decay process $b \to cl\bar{\nu}_l$, and can be uniformly described by the electroweak effective Hamiltonian in SM. However, these decay processes involve both electroweak and strong interactions, and are difficult to study with the perturbative filed theory. As one of powerful non-perturbative approaches to deal with hadronic parameters, the QCDSR is time-honored and widely used in studying the properties of hadrons~\cite{Shifman:1978bx,Shifman:1978by,Colangelo:2000dp,Wang:2025sic}. Specifically, the QCDSR based on three-point correlation function is used to analyze the hadron weak or electromagnetic transition form factors~\cite{Shi:2019hbf,Zhao:2021sje,Zhang:2023nxl,Lu:2024tgy,Lu:2025bvi,Lu:2025usr} and the coupling constants of strong vertices~\cite{Bracco:2011pg,Lu:2023pcg,Lu:2025zaf}. In our previous works, the form factors related to $\Lambda_b\to\Lambda_c$, $\Xi_b\to\Xi_c$, $\Xi_{cc}\to\Sigma^*_c$, $\Xi_{cc}\to\Xi'^*_c$, $\Omega_{cc}\to\Xi'^*_c$ and $\Omega_{cc}\to\Omega^*_c$ transitions are analyzed by three-point QCDSR, where all possible coupling of interpolating current to hadronic states are considered in phenomenological side and the Dirac structures independent form factors are obtained~\cite{Lu:2025gol,Yu:2026tbk}. As continuation and expansion of our previous works, we analyze the vector and axial vector form factors of $\Sigma_b\to\Sigma_c$, $\Xi'_b\to\Xi'_c$ and $\Omega_b\to\Omega_c$ by using three-point QCDSR in the present work. With obtained form factors, the corresponding semileptonic decay processes are also analyzed.  

This article is organized as follows. After introduction in Sec.~\ref{sec1}, the semileptonic decay processes $\Sigma_b\to\Sigma_cl\bar{\nu}_l$, $\Xi'_b\to\Xi'_cl\bar{\nu}_l$ and $\Omega_b\to\Omega_cl\bar{\nu}_l$ are analyzed in Sec.~\ref{sec2}, and the electroweak transition form factors are introduced. Then, these form factors are analyzed with in the framework of three-point QCDSR in Sec.~\ref{sec3}. Sec.~\ref{sec4} is devoted to
present the numerical results and discussions and Sec.~\ref{sec5} is
the conclusion part. Some important figures and tables are shown in Appendices~\ref{Sec:AppA} and \ref{Sec:AppB}.

\section{Semileptonic decays of $\Sigma_b\to\Sigma_cl\bar{\nu}_l$, $\Xi'_b\to\Xi'_cl\bar{\nu}_l$ and $\Omega_b\to\Omega_cl\bar{\nu}_l$}\label{sec2}
\begin{figure}
	\centering
	\includegraphics[width=8.5cm]{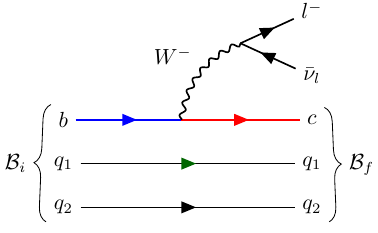}
	\caption{The Feynman diagram for semileptonic decays $\Sigma_b\to\Sigma_cl\bar{\nu}_l$, $\Xi'_b\to\Xi'_cl\bar{\nu}_l$ and $\Omega_b\to\Omega_cl\bar{\nu}_l$.}
	\label{FDH}
\end{figure}

The semileptonic decay processes $\Sigma_b\to\Sigma_cl\bar{\nu}_l$, $\Xi'_b\to\Xi'_cl\bar{\nu}_l$ and $\Omega_b\to\Omega_cl\bar{\nu}_l$ are all dominated by the transition $b\to cl\bar{\nu}_l$ at the quark level. The corresponding effective Hamiltonian can be written as the following form: 
\begin{eqnarray}\label{eq:1}
	H_{eff} = \frac{G_F}{\sqrt 2 }V_{cb}\bar c\gamma _\mu (1 - \gamma _5)b\bar{ v}_l\gamma _\mu (1 - \gamma _5)l,
\end{eqnarray}
where $G_F$ and $V_{cb}$ are the Fermi constant and CKM matrix element, respectively. With above Hamiltonian, the transition matrix element of these decays can be expressed as
\begin{eqnarray}\label{eq:2}
\notag
T &&= \left\langle \mathcal{B}_f(p')l(l)\bar{\nu}_l(v) \right|H_{eff}\left| \mathcal{B}_i(p) \right\rangle \\
\notag
&&= \frac{G_F}{\sqrt 2}V_{cb}\left\langle \mathcal{B}_f(p') \right|\bar c\gamma _\mu (1 - \gamma _5)b\left| \mathcal{B}_i(p) \right\rangle \\
&&\times \left\langle l(l)\bar{\nu}_l(v) \right|\bar{\nu}_l\gamma _\mu (1 - \gamma _5)l\left| 0 \right\rangle, 
\end{eqnarray}
where $\mathcal{B}_i$ and $\mathcal{B}_f$ denote the initial and final baryon states, respectively. The assignments of the initial and final state baryons and their quark compositions and quantum numbers for all decay processes are presented in Table~\ref{AB}. The corresponding Feynman diagram is shown as Fig.~\ref{FDH}.
\begin{table}[htbp]
	\begin{ruledtabular}
	\caption{The assignments of the initial and final state baryons and their quark compositions and quantum numbers.}
	\label{AB}\renewcommand\arraystretch{1.3}
	\begin{tabular}{c|c|c|c}
		Decay process&$\Sigma^-_b\to\Sigma^0_cl^-\bar{\nu}_l$&$\Xi'^-_b\to\Xi'^0_cl^-\bar{\nu}_l$&$\Omega^-_b\to\Omega^0_cl^-\bar{\nu}_l$ \\ \hline
		$\mathcal{B}_i$&$\Sigma^-_b$&$\Xi'^{-}_b$&$\Omega^-_b$ \\
		$\{q_1q_2\}Q$&$\{dd\}b$&$\{ds\}b$&$\{ss\}b$ \\ 
		$J^P$&$\frac{1}{2}^+$&$\frac{1}{2}^+$&$\frac{1}{2}^+$ \\ \hline
		$\mathcal{B}_f$&$\Sigma^0_c$&$\Xi'^{0}_c$&$\Omega^0_c$ \\
		$\{q_1q_2\}Q$&$\{dd\}c$&$\{ds\}c$&$\{ss\}c$ \\ 
		$J^P$&$\frac{1}{2}^+$&$\frac{1}{2}^+$&$\frac{1}{2}^+$ \\
	\end{tabular}
		\end{ruledtabular}
\end{table}

The leptonic part of the matrix element in Eq.~(\ref{eq:2}) can be expressed as the
following form by electroweak perturbation theory
\begin{eqnarray}\label{eq:3}
\langle l(l)\bar{\nu}_l(v) |\bar v_l\gamma _\mu(1 - \gamma _5)l | 0 \rangle  = \bar u_{v,s}\gamma _\mu(1 - \gamma _5)u_{-l,s'},
\end{eqnarray}
where $u_{v,s}$ and $u_{-l,s'}$ are the spinor wave functions for $\bar{\nu}_l$ and $l$, and the subscripts $v(l)$ and $s(s')$ denote the momentum and spin.

The hadronic part in Eq.~(\ref{eq:2}) can not be calculated by perturbation approach because QCD is non-perturbative at low energy region. However, according to Lorentz invariance and quantum number conservation, this matrix element can be expressed in terms of the following electroweak transition form factors,
\begin{eqnarray}\label{eq:4}
\notag
&&\left\langle \mathcal{B}_f(p') \right|\bar c\gamma _\mu(1 - \gamma _5)b \left| \mathcal{\mathcal{B}}_i(p) \right\rangle \\
\notag
&&= \bar u(p',s')\left[ \gamma _\mu f_1(q^2) + i\frac{\sigma _{\mu \nu }q^\nu }{m_{\mathcal{B}_i}}f_2(q^2) + \frac{q_\mu }{m_{\mathcal{B}_i}}f_3(q^2) \right]U(p,s)\\
\notag
&&- \bar u(p',s')\left[ \gamma _\mu g_1(q^2) + i\frac{\sigma _{\mu \nu }q^\nu }{m_{\mathcal{B}_i}}g_2(q^2) + \frac{q_\mu}{m_{\mathcal{B}_i}}g_3(q^2) \right]\gamma _5U(p,s),\\
\end{eqnarray}
where $q=p-p'$ denotes the transition momentum, $\sigma_{\mu\nu}=\frac{i}{2}[\gamma_\mu,\gamma_\nu]$. $U(p,s)$ and $u(p',s')$ are spinor wave functions of initial and final state baryons and $f_i(q^2)$ and $g_i(q^2)$ ($i=1,2,3$) represent the vector and axial vector transition form factors, respectively.

With Eqs.~(\ref{eq:2})-(\ref{eq:4}), the differential decay width for semileptonic decay $\mathcal{B}_i\to\mathcal{B}_f l\bar{\nu}_l$ can be expressed as the following forms by the helicity amplitudes,
\begin{eqnarray}\label{eq:5}
\notag
\frac{d\Gamma}{dq^2} &&= \frac{d\Gamma _L}{dq^2} + \frac{d\Gamma _T}{dq^2},\\
\notag
\frac{d\Gamma _L}{dq^2} &&= \frac{G_F^2V_{cb}^2q^2}{384\pi ^3m_{\mathcal{B}_i}^2}\frac{\sqrt {Q_+Q_-}}{2m_{\mathcal{B}_i}}{\left( 1 - \frac{m_l^2}{q^2} \right)^2}\\
\notag
&&\times \left[ \left(2 + \frac{m_l^2}{q^2}\right)\left(|H_{-\frac{1}{2},0}|^2 + |H_{\frac{1}{2},0}|^2\right) \right.\\
\notag
&&\left. { + \frac{3m_l^2}{q^2}\left(|H_{-\frac{1}{2},t}|^2 + |H_{\frac{1}{2},t}|^2\right)} \right],\\
\notag
\frac{d\Gamma _T}{dq^2} &&= \frac{G_F^2V_{cb}^2q^2}{384\pi ^3m_{\mathcal{B}_i}^2}\frac{\sqrt{Q_+Q_-}}{2m_{\mathcal{B}_i}}\left( 1 - \frac{m_l^2}{q^2} \right)^2\left( 2 + \frac{m_l^2}{q^2} \right)\\
&&\times \left(|H_{\frac{1}{2},1}|^2 + |H_{-\frac{1}{2}, - 1}|^2\right),
\end{eqnarray}
where $\Gamma_L$ and $\Gamma_T$ denote the longitudinally and transversely
polarized decay widths, $Q_{\pm}=(m_{\mathcal{B}_i}\pm m_{\mathcal{B}_f})^2-q^2$. The total helicity amplitudes can be written as:
\begin{eqnarray}\label{eq:6}
	H_{\lambda _f, \lambda _W}= H^V_{\lambda _f, \lambda _W}-H^A_{\lambda _f, \lambda _W}.
\end{eqnarray}
Here $\lambda_f$ and $\lambda_W$ represent the polarization of final state baryon and $W$ boson, respectively. The superscripts $V$ and $A$ denote the vector and axial vector helicity amplitudes. The full expressions of positive helicity amplitudes are given as follows,
\begin{eqnarray}\label{eq:7}
\notag
H_{\frac{1}{2},0}^V &&=  - i\frac{\sqrt {Q_-}}{\sqrt{q^2}}\left[ (m_{\mathcal{B}_i} + m_{\mathcal{B}_f})f_1(q^2) - \frac{q^2}{m_{\mathcal{B}_i}}f_2(q^2) \right],\\
\notag
H_{\frac{1}{2},1}^V &&= i\sqrt {2Q_-} \left[  - f_1(q^2) + \frac{m_{\mathcal{B}_i} + m_{\mathcal{B}_f}}{m_{\mathcal{B}_i}}f_2(q^2) \right],\\
\notag
H_{\frac{1}{2},t}^V &&=  - i\frac{\sqrt {Q_ +}}{\sqrt{q^2}}\left[ (m_{\mathcal{B}_i} - m_{\mathcal{B}_f})f_1(q^2) + \frac{q^2}{m_{\mathcal{B}_i}}f_3(q^2) \right],\\
\notag
H_{\frac{1}{2},0}^A &&=  - i\frac{\sqrt {Q_+}}{\sqrt{q^2}}\left[ (m_{\mathcal{B}_i} - m_{\mathcal{B}_f})g_1(q^2) + \frac{q^2}{m_{\mathcal{B}_i}}g_2(q^2) \right],\\
\notag
H_{\frac{1}{2},1}^A &&= i\sqrt {2Q_+} \left[- g_1(q^2) - \frac{m_{\mathcal{B}_i} - m_{\mathcal{B}_f}}{m_{\mathcal{B}_i}}g_2(q^2) \right],\\
H_{\frac{1}{2},t}^A &&=  - i\frac{\sqrt{Q_-}}{\sqrt{q^2}}\left[ (m_{\mathcal{B}_i} + m_{\mathcal{B}_f})g_1(q^2) - \frac{q^2}{m_{\mathcal{B}_i}}g_3(q^2) \right],
\end{eqnarray}
the relations of positive and negative helicity amplitudes are as follows,
\begin{eqnarray}\label{eq:8}
\notag
H_{-\lambda _f, -\lambda _W}^V = H_{\lambda _f,\lambda _W}^V,\\
H_{-\lambda _f, -\lambda _W}^A =  - H_{\lambda _f,\lambda _W}^A.
\end{eqnarray}
Finally, the total decay widths of these semileptonic decays can be obtained by integrating out the square momentum $q^2$.
\begin{eqnarray}\label{eq:9}
\Gamma=\Gamma_L+\Gamma_T=\int\limits_{m_l^2}^{(m_{\mathcal{B}_i}-m_{\mathcal{B}_f})^2}\frac{d\Gamma}{dq^2} dq^2.
\end{eqnarray}
\section{The QCD sum rules for transition form factors}\label{sec3}
To obtained the electroweak transition form factors, the following three-point correlation function is firstly constructed,
\begin{eqnarray}\label{eq:10}
\notag
\Pi _\mu (p',q) &&= i^2\int d^4x d^4ye^{ip'\cdot x}e^{iq\cdot y}\\
&&\times \left\langle 0 \right|\mathcal{T}[{J_{{\mathcal{B}_f}}}(x)J_\mu ^{V - A}(y){{\bar J}_{{\mathcal{B}_i}}}(0)]\left| 0 \right\rangle,
\end{eqnarray}
where $\mathcal{T}$ is the time ordered operator and $\bar{J}=J^\dagger\gamma_0$. $J_{\mathcal{B}_i}$ and $J_{\mathcal{B}_f}$ are the interpolating currents of initial and final state baryons, and $J^{V-A}$ is the electroweak transition current. These currents have the following forms,
\begin{eqnarray}\label{eq:11}
\notag
J_{\mathcal{B}_f}(x) &&= \varepsilon_{ijk}\left( q_1^{iT}(x)\mathcal{C}\gamma _\alpha q_2^j(x) \right)\gamma _\alpha \gamma _5 c^k(x),\\
\notag
J_\mu ^{V - A}(y) &&= \bar c^m(y)\gamma _\mu (1 - \gamma _5)b^m(y),\\
J_{\mathcal{B}_i}(0) &&= \varepsilon _{i'j'k'}\left( q_1^{i'T}(0)\mathcal{C}\gamma _\beta q_2^{j'}(0) \right)\gamma _\beta \gamma _5 b^{k'}(0),
\end{eqnarray}
where $\varepsilon_{ijk}$ is the 3 dimension Levi-Civita tensor, $i(i')$, $j(j')$, $k(k')$ and $m$ are the color indices and $\mathcal{C}$ represents the charge conjugation operator. $q_1$ and $q_2$ denote the light quarks in the initial and final state baryons which are listed in Table~\ref{AB}.

In the framework of QCDSR, the above correlation function can be calculated at both hadron and quark levels, where the former is called as phenomenological side and the latter is called as QCD side. By matching the calculations of these two levels and using the quark-hadron duality condition, the sum rule equations for the form factors
can be obtained.
\subsection{The phenomenological side}
The phenomenological treatment for the three-point correlation function begins by inserting the complete set of hadronic states coupled to the interpolating currents. After performing the coordinate space integrals and separating the ground state contribution from the excited states, we apply the double dispersion relation, which allows the three-point correlation function to be represented as the following form~\cite{Colangelo:2000dp},
\begin{eqnarray}\label{eq:12}
\notag
&&\Pi _\mu ^{\mathrm{phy}}(p^2,p'^2)\\
\notag
&&= \frac{\left\langle 0 \right|J_{\mathcal{B}_f} \left| \mathcal{B}_f^+(p') \right\rangle \left\langle \mathcal{B}_f^+(p') \right|J_\mu ^{V - A}\left| \mathcal{B}_i^+(p)\right\rangle \left\langle \mathcal{B}_i^+(p) \right|\bar J_{\mathcal{B}_i} \left| 0 \right\rangle}{(m_{\mathcal{B}_f^+}^2-p'^2)(m_{\mathcal{B}_i^+}^2-p^2)}\\
\notag
&&+ \frac{\left\langle 0 \right|J_{\mathcal{B}_f} \left| \mathcal{B}_f^-(p') \right\rangle \left\langle \mathcal{B}_f^-(p') \right|J_\mu ^{V - A}\left| \mathcal{B}_i^+(p)\right\rangle \left\langle \mathcal{B}_i^+(p) \right|\bar J_{\mathcal{B}_i} \left| 0 \right\rangle}{(m_{\mathcal{B}_f^-}^2-p'^2)(m_{\mathcal{B}_i^+}^2-p^2)}\\
\notag
&&+ \frac{\left\langle 0 \right|J_{\mathcal{B}_f} \left| \mathcal{B}_f^+(p') \right\rangle \left\langle \mathcal{B}_f^+(p') \right|J_\mu ^{V - A}\left| \mathcal{B}_i^-(p)\right\rangle \left\langle \mathcal{B}_i^-(p) \right|\bar J_{\mathcal{B}_i} \left| 0 \right\rangle}{(m_{\mathcal{B}_f^+}^2-p'^2)(m_{\mathcal{B}_i^-}^2-p^2)}\\
\notag
&&+ \frac{\left\langle 0 \right|J_{\mathcal{B}_f} \left| \mathcal{B}_f^-(p') \right\rangle \left\langle \mathcal{B}_f^-(p') \right|J_\mu ^{V - A}\left| \mathcal{B}_i^-(p)\right\rangle \left\langle \mathcal{B}_i^-(p) \right|\bar J_{\mathcal{B}_i} \left| 0 \right\rangle}{(m_{\mathcal{B}_f^-}^2-p'^2)(m_{\mathcal{B}_i^-}^2-p^2)} \\
&&+ h.c.,
\end{eqnarray}
where the currents are all at $x=0$, $h.c.$ denotes the contributions from higher resonances and continuum states of hadrons. The interpolating currents in Eq.~(\ref{eq:11}) with spin parity $J^P=\frac{1}{2}^+$ can not only couple to $J^P=\frac{1}{2}^+$ baryons but also to $J^P=\frac{1}{2}^-$ baryons, because that we can multiply the interpolating currents with $J^P=\frac{1}{2}^+$ by the Dirac matrix $\gamma_5$ to change its parity. For example, we introduce $J_+$ to represent the interpolating current with $J^P=\frac{1}{2}^+$, and it satisfies the relation $\mathcal{P}J_+(t,\vec{x})\mathcal{P}^{-1}=J_+(t,-\vec{x})$ under the parity transformation. The coupling of this current with $J^P=\frac{1}{2}^+$ baryon can be defined as $\langle0|J_+(0)|\mathcal{B}_+\rangle=\lambda_{\mathcal{B}_+}u_{\mathcal{B}_+}$. Multiplying $J_+$ by the Dirac matrix $\gamma_5$, the relation of current under parity transformation becomes $\mathcal{P}\gamma_5J_+(t,\vec{x})\mathcal{P}^{-1}=-\gamma_5J_+(t,-\vec{x})$. By introducing $J_-$ to represent the current with $J^P=\frac{1}{2}^-$, we have $J_-=\gamma_5J_+$. Since the coupling of $J_-$ with $J^P=\frac{1}{2}^-$ baryon can be defined as $\langle0|J_-(0)|\mathcal{B}_-\rangle=\lambda_{\mathcal{B}_-}u_{\mathcal{B}_-}$, the coupling of $J_+$ and $J^P=\frac{1}{2}^-$ baryon can be expressed as $\langle0|J_+(0)|\mathcal{B}_-\rangle=\lambda_{\mathcal{B}_-}\gamma_5u_{\mathcal{B}_-}$. In this analysis, all possible couplings of the interpolating current to hadronic states are considered. The corresponding hadron vacuum matrix elements can be written as, 
\begin{eqnarray}\label{eq:13}
\notag
\left\langle 0 \right|J_{\mathcal{B}_f}(0)\left|\mathcal{B}_f^+(p') \right\rangle  &&= \lambda _{\mathcal{B}_f^+}u(p',s'),\\
\notag
\left\langle 0 \right|J_{\mathcal{B}_f}(0)\left|\mathcal{B}_f^-(p') \right\rangle  &&= \lambda _{\mathcal{B}_f^-}\gamma_5u(p',s'),\\
\notag
\left\langle \mathcal{B}_i^+(p)\right|{\bar J}_{\mathcal{B}_i}(0)\left| 0 \right\rangle  &&= \lambda _{\mathcal{B}_i^+}\bar U(p,s),\\
\left\langle \mathcal{B}_i^-(p)\right|{\bar J}_{\mathcal{B}_i}(0)\left| 0 \right\rangle  &&=- \lambda _{\mathcal{B}_i^-}\bar U(p,s)\gamma_5,
\end{eqnarray}
here $\lambda_{\mathcal{B}_i}$ and $\lambda_{\mathcal{B}_f}$ are the pole residues of initial and final state baryons. For the convenience of calculation, the following simpler parameterization for form factors is employed in our present work
\begin{widetext}
\begin{eqnarray}\label{eq:14}
\notag
\left\langle \mathcal{B}_f^+(p') \right|J_\mu ^{V - A}(0)\left| \mathcal{B}_i^+(p) \right\rangle &&= \bar u(p',s')\left[F_1^{ +  + }(q^2)\frac{p^\mu }{m_{\mathcal{B}_i^+}} + F_2^{ +  + }(q^2)\frac{p'^\mu }{m_{\mathcal{B}_f^+}} + F_3^{ +  + }(q^2)\gamma _\mu \right]U(p,s)\\
\notag
&&- \bar u(p',s')\left[ G_1^{ +  + }(q^2)\frac{p^\mu }{m_{\mathcal{B}_i^+}} + G_2^{ +  + }(q^2)\frac{p'^\mu }{m_{\mathcal{B}_f^+}} + G_3^{ +  + }(q^2)\gamma _\mu \right]\gamma_5 U(p,s),\\
\notag
\left\langle \mathcal{B}_f^-(p') \right|J_\mu ^{V - A}(0)\left| \mathcal{B}_i^+(p) \right\rangle &&= \bar u(p',s')\gamma_5\left[  F_1^{ +  - }(q^2)\frac{p^\mu }{m_{\mathcal{B}_i^+}} + F_2^{ +  - }(q^2)\frac{p'^\mu }{m_{\mathcal{B}_f^-}} + F_3^{ +  - }(q^2)\gamma _\mu \right]U(p,s)\\
\notag
&&- \bar u(p',s')\gamma_5\left[   G_1^{ +  - }(q^2)\frac{p^\mu }{m_{\mathcal{B}_i^+}} + G_2^{ +  - }(q^2)\frac{p'^\mu }{m_{\mathcal{B}_f^-}} + G_3^{ +  - }(q^2)\gamma _\mu \right]\gamma_5 U(p,s),\\
\notag
\left\langle \mathcal{B}_f^+(p') \right|J_\mu ^{V - A}(0)\left| \mathcal{B}_i^-(p) \right\rangle &&= \bar u(p',s')\left[  F_1^{ -  + }(q^2)\frac{p^\mu }{m_{\mathcal{B}_i^-}} + F_2^{ -  + }(q^2)\frac{p'^\mu }{m_{\mathcal{B}_f^+}} + F_3^{ -  + }(q^2)\gamma _\mu \right]\gamma_5U(p,s)\\
\notag
&&- \bar u(p',s')\left[ G_1^{ -  + }(q^2)\frac{p^\mu }{m_{\mathcal{B}_i^-}} + G_2^{ -  + }(q^2)\frac{p'^\mu }{m_{\mathcal{B}_f^+}} + G_3^{ -  + }(q^2)\gamma _\mu \right]U(p,s),\\
\notag
\left\langle \mathcal{B}_f^-(p') \right|J_\mu ^{V - A}(0)\left| \mathcal{B}_i^-(p) \right\rangle  &&= \bar u(p',s')\gamma _5\left[ F_1^{ -  - }(q^2)\frac{p^\mu }{m_{\mathcal{B}_i^-}} + F_2^{ -  - }(q^2)\frac{p'^\mu }{m_{\mathcal{B}_f^-}} + F_3^{ -  - }(q^2)\gamma _\mu \right]\gamma _5U(p,s)\\
&&- \bar u(p',s')\gamma _5\left[  G_1^{ -  - }(q^2)\frac{p^\mu }{m_{\mathcal{B}_i^-}} + G_2^{ -  - }(q^2)\frac{p'^\mu }{m_{\mathcal{B}_f^-}} + G_3^{ -  - }(q^2)\gamma _\mu\right]U(p,s),
\end{eqnarray}
\end{widetext}
where the superscript of form factors denotes the parity of the initial and final state hadrons in the electroweak transition matrix elements. For example, $+-$ and $-+$ represent $\mathcal{B}_i^+\to\mathcal{B}_f^-$ and $\mathcal{B}_i^-\to\mathcal{B}_f^+$ transitions, respectively. From Eq.~(\ref{eq:14}), one can find that the twenty-four form factors will be introduced in the phenomenological side when all possible couplings of interpolating currents to hadronic states are considered. With Eqs.~(\ref{eq:12})-(\ref{eq:14}), the correlation function in the phenomenological side can be obtained as,
\begin{widetext}
\begin{eqnarray}\label{eq:15}
\notag
\Pi _\mu ^{\mathrm{phy}}(p^2,p'^2) &&= \frac{\lambda _{\mathcal{B}_f^+}\lambda _{\mathcal{B}_i^+}(\slashed p' + m_{\mathcal{B}_f^+})\left\{ \begin{array}{l}
\left[  F_1^{ +  + }(q^2)\frac{p^\mu }{m_{\mathcal{B}_i^+}} + F_2^{ +  + }(q^2)\frac{p'^\mu }{m_{\mathcal{B}_f^+}} + F_3^{ +  + }(q^2)\gamma _\mu \right]\\
- \left[  G_1^{ +  + }(q^2)\frac{p^\mu }{m_{\mathcal{B}_i^+}} + G_2^{ +  + }(q^2)\frac{p'^\mu }{m_{\mathcal{B}_f^+}} + G_3^{ +  + }(q^2)\gamma _\mu \right]\gamma _5
\end{array} \right\}(\slashed p +m_{\mathcal{B}_i^+})}{(m_{\mathcal{B}_f^+}^2-p'^2)(m_{\mathcal{B}_i^+}^2-p^2)}\\
\notag
&&+ \frac{\lambda _{\mathcal{B}_f^-}\lambda _{\mathcal{B}_i^+}\gamma _5(\slashed p' + m_{\mathcal{B}_f^-})\left\{ \begin{array}{l}
\gamma _5\left[  F_1^{ +  - }(q^2)\frac{p^\mu }{m_{\mathcal{B}_i^+}} + F_2^{ +  - }(q^2)\frac{p'^\mu }{m_{\mathcal{B}_f^-}} + F_3^{ +  - }(q^2)\gamma _\mu \right]\\
- \gamma _5\left[ G_1^{ +  - }(q^2)\frac{p^\mu }{m_{\mathcal{B}_i^+}} + G_2^{ +  - }(q^2)\frac{p'^\mu }{m_{\mathcal{B}_f^-}} + G_3^{ +  - }(q^2)\gamma _\mu \right]\gamma _5
\end{array} \right\}(\slashed p + m_{\mathcal{B}_i^+})}{(m_{\mathcal{B}_f^-}^2-p'^2)(m_{\mathcal{B}_i^+}^2-p^2)}\\
\notag
&&- \frac{\lambda _{\mathcal{B}_f^+}\lambda _{\mathcal{B}_i^-}(\slashed p' + m_{\mathcal{B}_f^+})\left\{ \begin{array}{l}
\left[ F_1^{ -  + }(q^2)\frac{p^\mu }{m_{\mathcal{B}_i^-}} + F_2^{ -  + }(q^2)\frac{p'^\mu }{m_{\mathcal{B}_f^+}} + F_3^{ -  + }(q^2)\gamma _\mu \right]\gamma _5\\
- \left[ G_1^{ -  + }(q^2)\frac{p^\mu }{m_{\mathcal{B}_i^-}} + G_2^{ -  + }(q^2)\frac{p'^\mu }{m_{\mathcal{B}_f^+}} + G_3^{ -  + }(q^2)\gamma _\mu \right]
\end{array} \right\}(\slashed p + m_{\mathcal{B}_i^-})\gamma _5}{(m_{\mathcal{B}_f^+}^2-p'^2)(m_{\mathcal{B}_i^-}^2-p^2)}\\
&&- \frac{\lambda _{\mathcal{B}_f^-}\lambda _{\mathcal{B}_i^-}\gamma _5(\slashed p' +m_{\mathcal{B}_f^-})\left\{ \begin{array}{l}
\gamma _5\left[ F_1^{ -  - }(q^2)\frac{p^\mu }{m_{\mathcal{B}_i^-}} + F_2^{ -  - }(q^2)\frac{p'^\mu }{m_{\mathcal{B}_f^-}} + F_3^{ -  - }(q^2)\gamma _\mu \right]\gamma _5\\
- \gamma _5\left[ G_1^{ -  - }(q^2)\frac{p^\mu }{m_{\mathcal{B}_i^-}} + G_2^{ -  - }(q^2)\frac{p'^\mu }{m_{\mathcal{B}_f^-}} + G_3^{ -  - }(q^2)\gamma _\mu \right]
\end{array} \right\}(\slashed p + m_{\mathcal{B}_i^-})\gamma _5}{(m_{\mathcal{B}_f^-}^2-p'^2)(m_{\mathcal{B}_i^-}^2-p^2)} +h.c.
\end{eqnarray}
\end{widetext}
The above correlation function can be decomposed into the following twenty-four independent dirac structures,
\begin{eqnarray}\label{eq:16}
\notag
\Pi^{\mathrm{phy}}_\mu(p^2,p'^2)&&=\Pi^{\mathrm{phy}}_1\gamma_\mu+\Pi^{\mathrm{phy}}_2\gamma_\mu\slashed p'+\Pi^{\mathrm{phy}}_3\gamma_\mu\slashed q + \Pi^{\mathrm{phy}}_4\gamma_\mu\slashed p'\slashed q\\
\notag
&&+\Pi^{\mathrm{phy}}_5\slashed p'p'_\mu+\Pi^{\mathrm{phy}}_6\slashed p'q_\mu+\Pi^{\mathrm{phy}}_7\slashed qp'_\mu+\Pi^{\mathrm{phy}}_8\slashed qq_\mu\\
\notag
&&+\Pi^{\mathrm{phy}}_9\slashed p'\slashed qp'_\mu+\Pi^{\mathrm{phy}}_{10}\slashed p'\slashed qq_\mu+\Pi^{\mathrm{phy}}_{11}p'_\mu+\Pi^{\mathrm{phy}}_{12}q_\mu\\
\notag
&&+\Pi^{\mathrm{phy}}_{13}\gamma_\mu\gamma_5+\Pi^{\mathrm{phy}}_{14}\gamma_\mu\slashed p'\gamma_5+\Pi^{\mathrm{phy}}_{15}\gamma_\mu\slashed q\gamma_5\\
\notag
&& + \Pi^{\mathrm{phy}}_{16}\gamma_\mu\slashed p'\slashed q\gamma_5+\Pi^{\mathrm{phy}}_{17}\slashed p'\gamma_5p'_\mu+\Pi^{\mathrm{phy}}_{18}\slashed p'\gamma_5q_\mu\\
\notag
&&+\Pi^{\mathrm{phy}}_{19}\slashed q\gamma_5p'_\mu+\Pi^{\mathrm{phy}}_{20}\slashed q\gamma_5q_\mu+\Pi^{\mathrm{phy}}_{21}\slashed p'\slashed q\gamma_5p'_\mu\\
&&+\Pi^{\mathrm{phy}}_{22}\slashed p'\slashed q\gamma_5q_\mu+\Pi^{\mathrm{phy}}_{23}\gamma_5p'_\mu+\Pi^{\mathrm{phy}}_{24}\gamma_5q_\mu.
\end{eqnarray}
The form factors in Eq.~(\ref{eq:15}) are included in the expansion coefficients $\Pi^{\mathrm{phy}}_i$ ($i=1,...,24$), which are commonly called as the scalar invariant amplitude. 

\subsection{The QCD side}
In QCD side, we substitute the explicit forms of the interpolating and transition currents given in Eq.~(\ref{eq:11}) into the correlation function defined in Eq.~(\ref{eq:10}). Carrying out the operator product expansion (OPE) via Wick's theorem yields the following expression for the correlation function:
\begin{eqnarray}\label{eq:17}
\notag
\Pi _\mu ^{\mathrm{QCD}}(p',q) &&=A\varepsilon _{ijk}\varepsilon _{i'j'k'}\int d^4x d^4ye^{ip'\cdot x}e^{iq\cdot y}\\
\notag
&&\times Tr\left\{ S_{q_2}^{k'j}(x)\gamma _\beta \mathcal{C}S_{q_1}^{ki'T}(x)\mathcal{C} \gamma _\alpha \right\}\gamma_\alpha\gamma_5\\
&&\times S_c^{j'm}(x - y)\gamma _\mu (1 - \gamma _5)S_b^{mi}(y)\gamma_5\gamma_\beta,
\end{eqnarray}
where $A=1$ is for $q_1\neq q_2$ and $A=2$ is for $q_1=q_2$. $S_{q_{1,2}}^{ij}(x)$, and $S_{c,b}^{ij}(x)$ are the full propagator of light and heavy quarks which can be written as follows~\cite{Pascual:1984zb,Reinders:1984sr},
\begin{eqnarray}\label{eq:18}
\notag
S_d^{ij}(x) &&= \frac{i}{(2\pi )^4}\int d^4k e^{ - ik\cdot x} \left\{ \frac{\delta ^{ij}}{\slashed k} - \frac{g_sG_{\alpha \beta }^nt_{ij}^n}{4}\frac{\sigma ^{\alpha \beta }\slashed k + \slashed k \sigma ^{\alpha \beta }}{(k^2)^2} \right.\\
\notag
&&-\left. \frac{g_s^2(t^at^b)_{ij}G_{\alpha \beta }^aG_{\mu \nu }^b(f^{\alpha \beta \mu \nu } + f^{\alpha \mu \beta \nu} + f^{\alpha \mu \nu \beta })}{4(k^2)^5} \right\}\\
\notag
&&- \frac{\delta ^{ij}\left\langle \bar qq \right\rangle}{12} - \frac{\delta ^{ij}x^2\left\langle \bar qg_s\sigma Gq \right\rangle }{192}- \frac{\left\langle \bar q^j\sigma ^{\mu \nu }q^i \right\rangle \sigma _{\mu \nu }}{8} \\
&&-\frac{\delta^{ij}x^4\left\langle \bar qq \right\rangle \left\langle g_s^2 GG \right\rangle}{27648}+ ...,
\end{eqnarray}
\begin{eqnarray}\label{eq:19}
\notag
S_s^{ij}(x) &&=\frac{i}{(2\pi )^4}\int d^4k e^{ - ik\cdot x} \left\{ \frac{\delta ^{ij}}{\slashed k-m_s} \right.\\
\notag
&&- \frac{g_sG_{\alpha \beta }^nt_{ij}^n}{4}\frac{\sigma ^{\alpha \beta }(\slashed k + m_s) + (\slashed k + m_s)\sigma ^{\alpha \beta }}{(k^2 - m_s^2)^2}\\
\notag
&&-\left. \frac{g_s^2(t^at^b)_{ij}G_{\alpha \beta }^aG_{\mu \nu }^b(f^{\alpha \beta \mu \nu } + f^{\alpha \mu \beta \nu} + f^{\alpha \mu \nu \beta })}{4(k^2 - m_s^2)^5} \right\}\\
\notag
&& - \frac{\delta ^{ij}\left\langle \bar ss \right\rangle }{12} + \frac{i\delta ^{ij}\slashed xm_s\left\langle \bar ss \right\rangle }{48}- \frac{\delta ^{ij}x^2\left\langle \bar sg_s\sigma Gs \right\rangle }{192}\\
\notag
&& + \frac{i\delta ^{ij}x^2\slashed xm_s\left\langle \bar s{g_s}\sigma Gs \right\rangle }{1152}- \frac{\left\langle {\bar s^j\sigma ^{\mu \nu }s^i} \right\rangle \sigma _{\mu \nu }}{8}\\
&&-\frac{\delta^{ij}x^4\left\langle \bar ss \right\rangle \left\langle g_s^2 GG \right\rangle}{27648}+...,
\end{eqnarray}
\begin{eqnarray}\label{eq:20}
\notag
S_{c[b]}^{ij}(x) &&= \frac{i}{(2\pi )^4}\int d^4k e^{ - ik\cdot x} \left\{ \frac{\delta ^{ij}}{\slashed k - m_{c[b]}} \right.\\
\notag
&&- \frac{g_sG_{\alpha \beta }^nt_{ij}^n}{4}\frac{\sigma ^{\alpha \beta }(\slashed k + m_{c[b]}) + (\slashed k + m_{c[b]})\sigma ^{\alpha \beta }}{(k^2 - m_{c[b]}^2)^2}\\
\notag
&&\left. - \frac{g_s^2(t^at^b)_{ij}G_{\alpha \beta }^aG_{\mu \nu }^b(f^{\alpha \beta \mu \nu } + f^{\alpha \mu \beta \nu} + f^{\alpha \mu \nu \beta })}{4(k^2 - m_{c[b]}^2)^5}+... \right\}.\\
\end{eqnarray}
Here $\langle\bar{q}(\bar{s})g_s\sigma Gq(s)\rangle=\langle\bar{q}(\bar{s})g_s\sigma_{\mu\nu} G^{\alpha}_{\mu\nu}t^{\alpha}q(s)\rangle$, $t^{\alpha}=\frac{\lambda^{\alpha}}{2}$, $\lambda^{\alpha}$ ($\alpha=1,...,8$) are the Gell-Mann matrices. $f^{\alpha\beta\mu\nu}$ has the following form,
\begin{eqnarray}\label{eq:21}
	\notag
	f^{\alpha \beta \mu \nu } &&= (\slashed k + m)\gamma ^\alpha (\slashed k + m)\gamma ^\beta(\slashed k + m)\gamma ^\mu(\slashed k + m)\gamma ^\nu(\slashed k + m),\\
\end{eqnarray}
where $m=m_{q_{1,2}}=m_d=0$, $m_s$ and $m_{c[b]}$ correspond to Eqs.~(\ref{eq:18}), (\ref{eq:19}) and (\ref{eq:20}), respectively.

\begin{figure*}
	\centering
	\includegraphics[width=18cm]{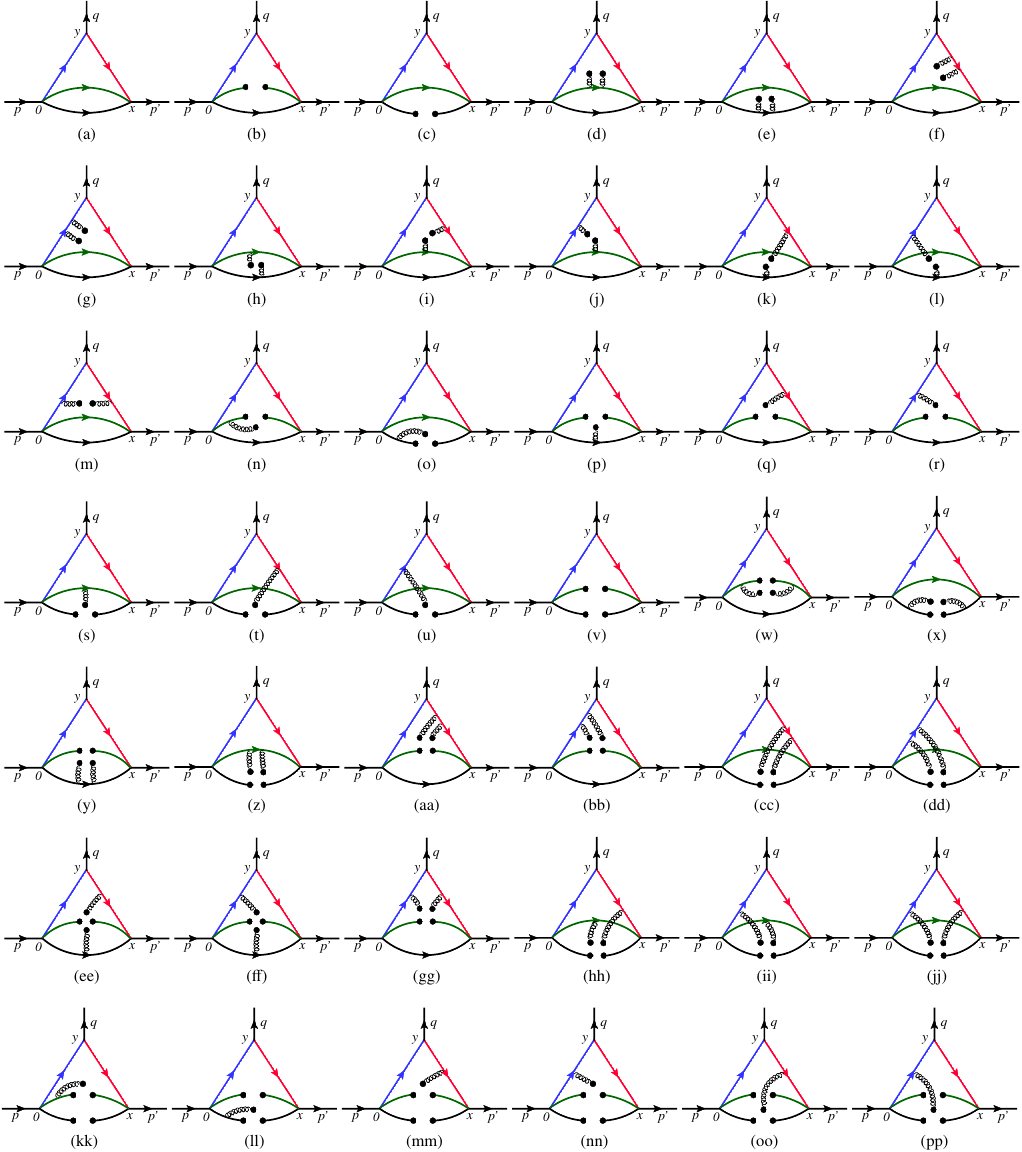}
	\caption{The Feynman diagrams for the perturbative part and vacuum condensate terms in quark level, where the blue, red, green and black solid lines denote the $b$, $c$, $q_1$ and $q_2$ quark lines, respectively. The black loop lines are the gluon lines.}
	\label{FDQ}
\end{figure*}

After rigorous calculation, the correlation function in QCD side can also be decomposed into the following twenty-four dirac structures:
\begin{eqnarray}\label{eq:22}
\notag
\Pi^{\mathrm{QCD}}_{\mu}(p^2,p'^2)&&=\Pi^{\mathrm{QCD}}_1\gamma_\mu+\Pi^{\mathrm{QCD}}_2\gamma_\mu\slashed p'+\Pi^{\mathrm{QCD}}_3\gamma_\mu\slashed q \\
\notag
&&+ \Pi^{\mathrm{QCD}}_4\gamma_\mu\slashed p'\slashed q+\Pi^{\mathrm{QCD}}_5\slashed p'p'_\mu+\Pi^{\mathrm{QCD}}_6\slashed p'q_\mu\\
\notag
&&+\Pi^{\mathrm{QCD}}_7\slashed qp'_\mu+\Pi^{\mathrm{QCD}}_8\slashed qq_\mu+\Pi^{\mathrm{QCD}}_9\slashed p'\slashed qp'_\mu\\
\notag
&&+\Pi^{\mathrm{QCD}}_{10}\slashed p'\slashed qq_\mu+\Pi^{\mathrm{QCD}}_{11}p'_\mu+\Pi^{\mathrm{QCD}}_{12}q_\mu\\
\notag
&&+\Pi^{\mathrm{QCD}}_{13}\gamma_\mu\gamma_5+\Pi^{\mathrm{QCD}}_{14}\gamma_\mu\slashed p'\gamma_5+\Pi^{\mathrm{QCD}}_{15}\gamma_\mu\slashed q\gamma_5 \\
\notag
&&+ \Pi^{\mathrm{QCD}}_{16}\gamma_\mu\slashed p'\slashed q\gamma_5+\Pi^{\mathrm{QCD}}_{17}\slashed p'\gamma_5p'_\mu+\Pi^{\mathrm{QCD}}_{18}\slashed p'\gamma_5q_\mu\\
\notag
&&+\Pi^{\mathrm{QCD}}_{19}\slashed q\gamma_5p'_\mu+\Pi^{\mathrm{QCD}}_{20}\slashed q\gamma_5q_\mu+\Pi^{\mathrm{QCD}}_{21}\slashed p'\slashed q\gamma_5p'_\mu\\
\notag
&&+\Pi^{\mathrm{QCD}}_{22}\slashed p'\slashed q\gamma_5q_\mu+\Pi^{\mathrm{QCD}}_{23}\gamma_5p'_\mu+\Pi^{\mathrm{QCD}}_{24}\gamma_5q_\mu,\\
\end{eqnarray}
where $\Pi^{\mathrm{QCD}}_i$ represents the scalar invariant amplitude in QCD side, and can be expressed as the following form by double dispersion relation:
\begin{eqnarray}\label{eq:23}
\Pi _i ^{\mathrm{QCD}}(p^2,p'^2) = \int\limits_{{u_{\min }}}^\infty  du \int\limits_{{s_{\min }}}^\infty  ds \frac{\rho _i ^{\mathrm{QCD}}(s,u,q^2)}{(s - p^2)(u - p'^2)}.
\end{eqnarray}
Here $\rho^{\mathrm{QCD}}_i(s,u,q^2)$ is the spectral density with $s=p^2$ and $u=p'^2$. $s_{\min}$ and $u_{\min}$ are kinetic limit for initial and final state baryons and their values are usually taken as the square of the summation of the quark masses that make up the corresponding hadrons. The QCD spectral density can be represented as the summation of perturbative part and different vacuum condensate terms:
\begin{eqnarray}\label{eq:24}
\notag
\rho _i ^{\mathrm{QCD}}(s,u,q^2) &&=\rho_i^{\mathrm{pert}}(s,u,q^2)+\rho_i^{\langle \bar{q}q\rangle}(s,u,q^2)\\
\notag
&&+\rho_i^{\langle \bar{s}s\rangle}(s,u,q^2)+\rho_i^{\langle g^2_sGG\rangle}(s,u,q^2)\\
\notag
&&+\rho_i^{\langle \bar{q}g_s\sigma Gq\rangle}(s,u,q^2)+\rho_i^{\langle \bar{s}g_s\sigma Gs\rangle}(s,u,q^2)\\
\notag
&&+\rho_i^{\langle \bar{q}q\rangle^2}(s,u,q^2)+\rho_i^{\langle \bar{q}q\rangle \langle \bar{s}s\rangle}(s,u,q^2)\\
\notag
&&+\rho_i^{\langle \bar{s}s\rangle^2}(s,u,q^2)+\rho_i^{\langle \bar{q}q\rangle\langle g_s^2GG\rangle}(s,u,q^2)\\
\notag
&&+\rho_i^{\langle \bar{s}s\rangle\langle g_s^2GG\rangle}(s,u,q^2)+\rho_i^{\langle \bar{q}q\rangle\langle \bar{q}g_s\sigma Gq\rangle}(s,u,q^2)\\
\notag
&&+\rho_i^{\langle \bar{q}q\rangle\langle \bar{s}g_s\sigma Gs\rangle}(s,u,q^2)+\rho_i^{\langle \bar{s}s\rangle\langle \bar{q}g_s\sigma Gq\rangle}(s,u,q^2)\\
&&+\rho_i^{\langle \bar{s}s\rangle\langle \bar{s}g_s\sigma Gs\rangle}(s,u,q^2).
\end{eqnarray}
In the present work, the contributions from perturbative part and vacuum condensate terms up to dimension 8 are all considered, and the corresponding forty-two Feynman diagrams are shown in detailed in Fig.~\ref{FDQ}. 

For the perturbative part (Fig.~\ref{FDQ} (a)), we firstly substituted the free propagators in momentum space of both light and heavy quarks in Eq.~(\ref{eq:17}). By performing the integration in coordinate space, the perturbative part correlation function can be written as:
\begin{eqnarray}\label{eq:25}
\notag
\Pi_{0\mu}^{\mathrm{QCD}}(p,q) &&= \frac{6A}{(2\pi )^8}\int d^4k_1 d^4k_2d^4k_3d^4k_4\delta ^4(q+k_3-k_4)\\
\notag
&&\times \delta ^4(p'-k_1-k_2-k_3)Tr\{(\slashed k_1 + m_{q_2})\gamma _\beta(\slashed k_2 - m_{q_1}) \\
\notag
&&\times\gamma _\alpha\}\frac{{{\gamma _\alpha }{\gamma _5}({{\slashed k}_3} + {m_c}){\gamma _\mu }(1 - {\gamma _5})({{\slashed k}_4} + {m_b}){\gamma _5}{\gamma _\beta }}}{{(k_1^2 - m_{{q_2}}^2)(k_2^2 - m_{{q_1}}^2)(k_3^2 - m_c^2)(k_4^2 - m_b^2)}},\\
\end{eqnarray}
where the subscript zero denotes the dimension of perturbative term. By setting all quark lines on-shell with the Cutkosky's rule~\cite{Cutkosky:1960sp}, the QCD spectral density function of perturbative term can be obtained as:
\begin{eqnarray}\label{eq:26}
\notag
\rho _{0\mu }^{\mathrm{QCD}}(s,u,q^2) &&=-\frac{6A}{(2\pi )^8}\frac{(-2\pi i)^5}{(2\pi i)^3}\int\limits_{(m_{q_1} + m_{q_2})^2}^{(\sqrt u-m_c)^2} dr \int d^4k_1 d^4k_3\\
\notag
&&\times \delta(k_1^2-m_{q_2}^2)\delta[(q'-k_1)^2-m_{q_1}^2]\\
\notag
&&\times \delta [(p' - k_3)^2 - r]\delta [(k_3 + q)^2 - m_b^2]\delta (k_3^2 - m_c^2)\\
\notag
&&\times Tr\{ (\slashed k_1 + m_{q_2})\gamma _\beta(\slashed p' - \slashed k_3 - \slashed k_1 - m_{q_1})\gamma _\alpha \} \\
\notag
&&\times \gamma _\alpha \gamma _5(\slashed k_3 + m_c)\gamma _\mu (1 - \gamma _5)[\slashed q + \slashed k_3 + m_b]\gamma _5\gamma _\beta,\\
\end{eqnarray}
where $q'=k_1+k_2$ and $r=q'^2$. The integral formulas for two and three Dirac delta functions can be found in Ref.~\cite{Lu:2025zaf}. 

The quark condensate term $\langle\bar{q}q\rangle$ with dimension 3 comes from the full propagator of light quarks. Since the heavy quark will not contribute to this condensation, there are only two Feynman diagrams for this term (See Figs.~\ref{FDQ} (b) and (c)). Taking Fig.~\ref{FDQ} (b) as an example, the corresponding correlation function in QCD side can be obtained as,
\begin{eqnarray}\label{eq:27}
\notag
\Pi _{3b\mu}^{\mathrm{QCD}}(p,q) &&=\frac{A\left\langle \bar q_1q_1 \right\rangle i}{2(2\pi )^4}\int d^4k_3Tr\{\gamma _\beta(\slashed p'-\slashed k_3 - m_{q_2})\gamma _\alpha\}  \\
\notag
&&\times \frac{\gamma _\alpha \gamma _5(\slashed k_3 + m_c)\gamma _\mu (1 - \gamma _5)(\slashed k_3 + \slashed q + m_b)\gamma _5\gamma _\beta }{(k_3^2 - m_c^2)[(p' - k_3)^2 - m_{q_2}^2][(k_3 + q)^2 - m_b^2]}.\\
\end{eqnarray} 
The corresponding spectral density function can also be obtained as the following form by Cutkosky's rule,
\begin{eqnarray}\label{eq:28}
\notag
\rho _{3b\mu}^{\mathrm{QCD}}(s,u,q^2) &&= \frac{A\left\langle \bar q_1q_1 \right\rangle i}{2(2\pi )^4}\frac{(-2\pi i)^3}{(2\pi i)^2}\int d^4k_3\delta(k_3^2- m_c^2) \\
\notag
&&\times \delta[(p' - k_3)^2 - m_{q_2}^2]\delta[(k_3 + q)^2 - m_b^2]\\
\notag
&&\times Tr\{ \gamma _\beta(\slashed p' - \slashed k_3 - m_{q_2})\gamma _\alpha\} \\
\notag
&&\times\gamma _\alpha \gamma _5(\slashed k_3 + m_c)\gamma _\mu (1 - \gamma _5)(\slashed k_3 + \slashed q + m_b)\gamma _5\gamma _\beta.\\
\end{eqnarray} 

For the gluon condensate $\langle g_s^2GG\rangle$ with dimension 4, there are ten Feynman diagrams, which are shown in Figs.~\ref{FDQ} (d)-(m). Taking Fig.~\ref{FDQ} (e) as an example, the corresponding correlation function in QCD side can be expressed as,
\begin{eqnarray}\label{eq:29}
\notag
\Pi _{4n\mu}^{\mathrm{QCD}}(p,q) &&= \frac{A\langle g_s^2GG \rangle }{48(2\pi )^8}\int d^4k_1 d^4k_2d^4k_3d^4k_4\delta ^4(q+k_3-k_4)\\
\notag
&&\times \delta ^4(p'-k_1-k_2-k_3)(g_{\rho\lambda}g_{\sigma\chi} - g_{\rho\chi}g_{\sigma\lambda})\\
\notag
&&\times Tr\{ (f^{\rho \sigma \lambda \chi}+ f^{\rho \lambda \sigma \chi} + f^{\rho \lambda \chi \sigma})\gamma _\beta \\
\notag
&&\times(\slashed p' - \slashed k_3 - \slashed k_1- m_{q_2})\gamma _\alpha\}\\
&&\times\frac{\gamma _\alpha \gamma _5(\slashed k_3 + m_c)\gamma _\mu(1 -\gamma _5)(\slashed k_4 + m_b)\gamma _5\gamma _\beta}{(k_1^2 - m_{q_2}^2)^5(k_2^2 - m_{q_1}^2)(k_3^2 - m_c^2)(k_4^2 - m_b^2)}.
\end{eqnarray} 
After using the following derivative formula:
\begin{eqnarray}\label{eq:30}
\frac{1}{(k^2-m^2)^n}=\frac{1}{(n-1)!}\frac{\partial^{(n-1)}}{(\partial A)^{(n-1)}}\frac{1}{k^2-A} \big{|}_{A\to m^2},
\end{eqnarray} 
its spectral density function can be obtained using steps similar to those for the perturbative part. The calculation methods of the QCD spectral density functions for the vacuum condensates with other dimensions can be found in Refs.~\cite{Lu:2025gol,Yu:2026tbk,Lu:2025zaf}.
\subsection{QCD sum rules for form factors}
Taking the variables change $p^2\to-P^2$, $p'^2\to-P'^2$ and
$q^2\to-Q^2$, we can perform the double Borel transforms~\cite{Reinders:1984sr} for variables $P^2$ and $P'^2$ both in phenomenological and QCD sides which can further suppress the contribution from higher resonance and continuum states hadron in phenomenological side. Furthermore, it also suppresses the contribution from higher dimension vacuum condensates in QCD side and improve the convergence of OPE. After double Borel transformation, the variables $P^2$ and $P'^2$ will be replaced by Borel parameters $T_1^2$ and $T_2^2$. For simplicity, the relations $T^2=T_1^2$, $T_2^2=kT_1^2=kT^2$ and $k=\frac{m_{\mathcal{B}_f}^2}{m_{\mathcal{B}_i}^2}$~\cite{Bracco:2011pg} are introduced to reduce the Borel parameters in this work. Then, using the quark-hadron duality condition, we can establish a series of linear equations about twenty-four scalar invariant amplitudes in both phenomenological and QCD sides. Finally, all form factors in Eq.~(\ref{eq:15}) can be uniquely determined by solving these twenty-four linear equations. In this article, we focus only on the transition form factors from positive parity to positive parity. The QCD sum rules of these form factors can be expressed as follows,
\begin{widetext}
\begin{eqnarray}\label{eq:31}
\notag
F_1^{ +  + }(Q^2) &&= \frac{m_{\mathcal{B}_i^+}e^{m^2_{\mathcal{B}_i^+}/T^2 + m^2_{\mathcal{B}_f^+}/kT^2}}{\lambda _{\mathcal{B}_f^+}{\lambda _{\mathcal{B}_i^+}}(m_{\mathcal{B}_f^+} + m_{\mathcal{B}_f^-})(m_{\mathcal{B}_i^+} + m_{\mathcal{B}_i^-})}\int\limits_{{u_{\min }}}^{u_0} du \int\limits_{{s_{\min }}}^{s_0} ds e^{ - s/T^2 - u/kT^2}\\
\notag
&&\times \left\{ {\rho _{12}^{\mathrm{QCD}}(s,u,Q^2) + m_{\mathcal{B}_f^-}\rho _6^{\mathrm{QCD}}(s,u,Q^2) + (m_{\mathcal{B}_i^-} - m_{\mathcal{B}_f^+})\left[m_{\mathcal{B}_f^-}\rho _{10}^{\mathrm{QCD}}(s,u,Q^2) + \rho _8^{\mathrm{QCD}}(s,u,Q^2)\right]} \right\},
\end{eqnarray}
\begin{eqnarray}
\notag
F_2^{ +  + }(Q^2) &&= \frac{m_{\mathcal{B}_f^+}e^{m^2_{\mathcal{B}_i^+}/T^2 + m^2_{\mathcal{B}_f^+}/kT^2}}{\lambda _{\mathcal{B}_f^+}{\lambda _{\mathcal{B}_i^+}}(m_{\mathcal{B}_f^+} + m_{\mathcal{B}_f^-})(m_{\mathcal{B}_i^+} + m_{\mathcal{B}_i^-})}\int\limits_{{u_{\min }}}^{u_0} du \int\limits_{{s_{\min }}}^{s_0} ds e^{-s/T^2 - u/kT^2}\\
\notag
&&\times \left\{ \begin{array}{l}
\rho _{11}^{\mathrm{QCD}}(s,u,Q^2) - \rho _{12}^{\mathrm{QCD}}(s,u,Q^2) + 2\left[\rho _2^{\mathrm{QCD}}(s,u,Q^2) - \rho _3^{\mathrm{QCD}}(s,u,Q^2)\right]\\
+ m_{\mathcal{B}_f^-}\left[2\rho _4^{\mathrm{QCD}}(s,u,Q^2) + \rho _5^{\mathrm{QCD}}(s,u,Q^2) - \rho _6^{\mathrm{QCD}}(s,u,Q^2)\right]\\
+ (m_{\mathcal{B}_i^-} - m_{\mathcal{B}_f^+})\left[m_{\mathcal{B}_f^-}\rho _9^{\mathrm{QCD}}(s,u,Q^2) - m_{\mathcal{B}_f^-}\rho _{10}^{\mathrm{QCD}}(s,u,Q^2) + 2\rho _4^{\mathrm{QCD}}(s,u,Q^2) + \rho _7^{\mathrm{QCD}}(s,u,Q^2) \right.\\ \left. - \rho _8^{\mathrm{QCD}}(s,u,Q^2)\right]
\end{array} \right\},
\end{eqnarray}
\begin{eqnarray}
\notag
F_3^{ +  + }(Q^2) &&= \frac{e^{m^2_{\mathcal{B}_i^+}/T^2 + m^2_{\mathcal{B}_f^+}/kT^2}}{\lambda _{\mathcal{B}_f^+}{\lambda _{\mathcal{B}_i^+}}(m_{\mathcal{B}_f^+} + m_{\mathcal{B}_f^-})(m_{\mathcal{B}_i^+} + m_{\mathcal{B}_i^-})}\int\limits_{{u_{\min }}}^{u_0} du \int\limits_{{s_{\min }}}^{s_0} ds e^{-s/T^2 - u/kT^2}\\
\notag
&&\times \left\{ \rho _1^{\mathrm{QCD}}(s,u,Q^2) + (m_{\mathcal{B}_i^-} + m_{\mathcal{B}_f^+})\rho _3^{\mathrm{QCD}}(s,u,Q^2) - m_{\mathcal{B}_f^-}\left[\rho _2^{\mathrm{QCD}}(s,u,Q^2) + (m_{\mathcal{B}_i^-} + m_{\mathcal{B}_f^+})\rho _4^{\mathrm{QCD}}(s,u,Q^2)\right] \right\},\\
\end{eqnarray}
\begin{eqnarray}\label{eq:32}
\notag
G_1^{ +  + }(Q^2) &&= \frac{m_{\mathcal{B}_i^+}e^{m^2_{\mathcal{B}_i^+}/T^2 + m^2_{\mathcal{B}_f^+}/kT^2}}{\lambda _{\mathcal{B}_f^+}{\lambda _{\mathcal{B}_i^+}}(m_{\mathcal{B}_f^+} + m_{\mathcal{B}_f^-})(m_{\mathcal{B}_i^+} + m_{\mathcal{B}_i^-})}\int\limits_{{u_{\min }}}^{u_0} du \int\limits_{{s_{\min }}}^{s_0} ds e^{ - s/T^2 - u/kT^2}\\
\notag
&&\times \left\{ \rho _{24}^{\mathrm{QCD}}(s,u,Q^2) + m_{\mathcal{B}_f^-}\rho _{18}^{\mathrm{QCD}}(s,u,Q^2) - (m_{\mathcal{B}_i^-} + m_{\mathcal{B}_f^+})\left[m_{\mathcal{B}_f^-}\rho _{22}^{\mathrm{QCD}}(s,u,Q^2) + \rho _{20}^{\mathrm{QCD}}(s,u,Q^2)\right] \right\},
\end{eqnarray}
\begin{eqnarray}
\notag
G_2^{ +  + }(Q^2) &&= \frac{m_{\mathcal{B}_f^+}e^{m^2_{\mathcal{B}_i^+}/T^2 + m^2_{\mathcal{B}_f^+}/kT^2}}{\lambda _{\mathcal{B}_f^+}{\lambda _{\mathcal{B}_i^+}}(m_{\mathcal{B}_f^+} + m_{\mathcal{B}_f^-})(m_{\mathcal{B}_i^+} + m_{\mathcal{B}_i^-})}\int\limits_{{u_{\min }}}^{u_0} du \int\limits_{{s_{\min }}}^{s_0} ds e^{-s/T^2 - u/kT^2}\\
\notag
&&\times \left\{ \begin{array}{l}
\rho _{23}^{\mathrm{QCD}}(s,u,Q^2) - \rho _{24}^{\mathrm{QCD}}(s,u,Q^2) + 2\left[\rho _{14}^{\mathrm{QCD}}(s,u,Q^2) - \rho _{15}^{\mathrm{QCD}}(s,u,Q^2)\right]\\
+ m_{\mathcal{B}_f^-}\left[2\rho _{16}^{\mathrm{QCD}}(s,u,Q^2) + \rho _{17}^{\mathrm{QCD}}(s,u,Q^2) - \rho _{18}^{\mathrm{QCD}}(s,u,Q^2)\right]\\
+ (m_{\mathcal{B}_i^-} + m_{\mathcal{B}_f^+})\left[m_{\mathcal{B}_f^-}\rho _{22}^{\mathrm{QCD}}(s,u,Q^2) - m_{\mathcal{B}_f^-}\rho _{21}^{\mathrm{QCD}}(s,u,Q^2) - 2\rho _{16}^{\mathrm{QCD}}(s,u,Q^2) - \rho _{19}^{\mathrm{QCD}}(s,u,Q^2)\right.\\
+\left. \rho _{20}^{\mathrm{QCD}}(s,u,Q^2)\right]
\end{array} \right\},
\end{eqnarray}
\begin{eqnarray}
\notag
G_3^{ +  + }(Q^2) &&= \frac{e^{m^2_{\mathcal{B}_i^+}/T^2 + m^2_{\mathcal{B}_f^+}/kT^2}}{\lambda _{\mathcal{B}_f^+}{\lambda _{\mathcal{B}_i^+}}(m_{\mathcal{B}_f^+} + m_{\mathcal{B}_f^-})(m_{\mathcal{B}_i^+} + m_{\mathcal{B}_i^-})}\int\limits_{{u_{\min }}}^{u_0} du \int\limits_{{s_{\min }}}^{s_0} ds e^{ - s/T^2 - u/kT^2}\\
\notag
&&\times \left\{ {\rho _{13}^{\mathrm{QCD}}(s,u,Q^2) - (m_{\mathcal{B}_i^-} - m_{\mathcal{B}_f^+})\rho _{15}^{\mathrm{QCD}}(s,u,Q^2) - m_{\mathcal{B}_f^-}\left[\rho _{14}^{\mathrm{QCD}}(s,u,Q^2) - (m_{\mathcal{B}_i^-} - m_{\mathcal{B}_f^+})\rho _{16}^{\mathrm{QCD}}(s,u,Q^2)\right]} \right\}.\\
\end{eqnarray}
\end{widetext}
Here $s_0$ and $u_0$ are threshold parameters for initial and final state hadrons which are introduced to eliminate the contributions of higher resonances and continuum states. They commonly fulfill the relation $u_0(s_0)=(m_{\mathrm{ground}}+\Delta)^2$, where the $m_{\mathrm{ground}}$ denotes the mass of ground state hadron, and $\Delta$ is the energy gap between the ground
and first excited states, commonly taken as a value of $0.3-0.7$ GeV which is based on experimental data and previous QCDSR calculation. For convenience, we ignore the superscripts of the form factors below and use $F_i$ and $G_i$ to represent the $F_i^{++}$ and $G_i^{++}$.
\section{Numerical results and discussions}\label{sec4}
The input parameters used in this work are all collected in Table~\ref{IP}. The mass of $d$ quark is ignored in this analysis, and the masses of $b$, $c$ and $s$ quarks and the values of vacuum condensate have energy scale dependence and satisfy the following renormalization group equations:  
\begin{eqnarray}\label{eq:33}
	\notag
	m_{c[b]}(\mu ) &&= m_{c[b]}(m_{c[b]})\left[\frac{\alpha _s(\mu )}{\alpha _s(m_{c[b]})}\right]^{\frac{12}{33 - 2N_f}},\\
	\notag
	m_s(\mu ) &&= m_s(2 \mathrm{GeV})\left[\frac{\alpha _s(\mu )}{\alpha _s(2 \mathrm{GeV})}\right]^{\frac{12}{33 - 2N_f}},\\
	\notag
	\left\langle \bar qq \right\rangle (\mu ) &&= \left\langle \bar qq \right\rangle (1\mathrm{GeV})\left[\frac{\alpha _s(1\mathrm{GeV})}{\alpha _s(\mu )}\right]^{\frac{12}{33 - 2N_f}},\\
	\notag
	\left\langle \bar ss \right\rangle (\mu ) &&= \left\langle \bar ss \right\rangle (1\mathrm{GeV})\left[\frac{\alpha _s(1\mathrm{GeV})}{\alpha _s(\mu )}\right]^{\frac{12}{33 - 2N_f}},\\
	\notag
	\left\langle \bar qg_s\sigma Gq \right\rangle (\mu ) &&= \left\langle \bar qg_s\sigma Gq \right\rangle (1\mathrm{GeV})\left[\frac{\alpha _s(1\mathrm{GeV})}{\alpha _s(\mu )}\right]^{\frac{2}{33 - 2N_f}},\\
	\notag
	\left\langle \bar sg_s\sigma Gs \right\rangle (\mu ) &&= \left\langle \bar sg_s\sigma Gs \right\rangle (1\mathrm{GeV})\left[\frac{\alpha _s(1\mathrm{GeV})}{\alpha _s(\mu )}\right]^{\frac{2}{33 - 2N_f}},\\
\end{eqnarray}
with
\begin{eqnarray}\label{eq:34}
	\notag
	\alpha _s(\mu ) &&= \frac{1}{b_0t}\left[ 1 - \frac{b_1}{b_0^2}\frac{\log t}{t} + \frac{b_1^2(\log ^2t - \log t - 1) + b_0b_2}{b_0^4t^2} \right],\\
\end{eqnarray}
where $t=\log\left(\frac{\mu^2}{\Lambda_{\mathrm{QCD}}^2}\right)$, $b_0=\frac{33-2N_f}{12\pi}$, $b_1=\frac{153-19N_f}{24\pi^2}$ and $b_2=\frac{2857-\frac{5033}{9}N_f+\frac{325}{27}N_f^2}{128\pi^3}$. The value of $\Lambda_{\mathrm{QCD}}$ is taken as $213$ MeV for the quark flavors $N_f=5$~\cite{ParticleDataGroup:2024cfk}. The minimum subtraction masses of $b$, $c$ and $s$ quarks are taken from the PDG, which are $m_b(m_b)=4.18\pm0.03$ GeV, $m_c(m_c)=1.275\pm0.025$ GeV and $m_s(\mu=2 \mathrm{GeV})=0.095\pm0.005$ GeV~\cite{ParticleDataGroup:2024cfk}. The energy scales are taken as $\mu=2$, $2.1$ and $2.2$ GeV for $\Sigma_b$, $\Xi'_b$ and $\Omega_b$ transitions in this analysis, respectively.
\begin{table}[htbp]
\begin{ruledtabular}\caption{Input parameters (IP) in this work. The values of vacuum condensate are at the energy scale $\mu=1$ GeV.}
\label{IP}
\begin{tabular}{c c c c }
IP&Values(GeV) &IP&Values \\ \hline
$m_{\Sigma_b(\frac{1}{2}^+)}$&5.816~\cite{ParticleDataGroup:2024cfk}&$\lambda_{\Sigma_b(\frac{1}{2}^+)}$&$\sqrt{2}\times0.062$ GeV$^3$\footnote{Since the quark compositions for $\Sigma_b$ and $\Sigma_c$ in the present work are different from these in Ref.~\cite{Wang:2009cr}. The factor $\sqrt{2}$ is introduced to modified the pole residues.}~\cite{Wang:2009cr} \\
$m_{\Sigma_b(\frac{1}{2}^-)}$&6.107~\cite{Yu:2022ymb}&$\lambda_{\Xi'_b(\frac{1}{2}^+)}$&$0.079$ GeV$^3$~\cite{Wang:2009cr} \\
$m_{\Xi'_b(\frac{1}{2}^+)}$&5.935~\cite{ParticleDataGroup:2024cfk}&$\lambda_{\Omega_b(\frac{1}{2}^+)}$&$0.134$ GeV$^3$~\cite{Wang:2009cr} \\
$m_{\Xi'_b(\frac{1}{2}^-)}$&6.232~\cite{Li:2022xtj}&$\lambda_{\Sigma_c(\frac{1}{2}^+)}$&$\sqrt{2}\times0.045$ GeV$^3$~\cite{Wang:2009cr} \\
$m_{\Omega_b(\frac{1}{2}^+)}$&6.046~\cite{ParticleDataGroup:2024cfk}&$\lambda_{\Xi'_c(\frac{1}{2}^+)}$&$0.055$ GeV$^3$~\cite{Wang:2009cr}\\
$m_{\Omega_b(\frac{1}{2}^-)}$&6.329~\cite{Yu:2022ymb}&$\lambda_{\Omega_c(\frac{1}{2}^+)}$&$0.093$ GeV$^3$~\cite{Wang:2009cr} \\
$m_{\Sigma_c(\frac{1}{2}^+)}$&2.454~\cite{ParticleDataGroup:2024cfk}&$\langle \bar{q}q\rangle$&$-(0.24\pm0.01)^3$ GeV$^3$~\cite{Shifman:1978by,Reinders:1984sr} \\
$m_{\Sigma_c(\frac{1}{2}^-)}$&2.809~\cite{Yu:2022ymb}&$\langle \bar{s}s\rangle$&$(0.8\pm0.1)\langle\bar{q}q\rangle$~\cite{Shifman:1978by,Reinders:1984sr} \\
$m_{\Xi'_c(\frac{1}{2}^+)}$&2.579~\cite{ParticleDataGroup:2024cfk}&$\langle \bar{q}g_s\sigma Gq\rangle$&$m_0^2\langle\bar{q}q\rangle$~\cite{Shifman:1978by,Reinders:1984sr}\\
$m_{\Xi'_c(\frac{1}{2}^-)}$&2.941~\cite{Li:2022xtj}&$\langle \bar{s}g_s\sigma Gs\rangle$&$m_0^2\langle\bar{s}s\rangle$~\cite{Shifman:1978by,Reinders:1984sr} \\
$m_{\Omega_c(\frac{1}{2}^+)}$&2.695~\cite{ParticleDataGroup:2024cfk}&$m_0^2$&$0.8\pm0.1$ GeV$^2$~\cite{Shifman:1978by,Reinders:1984sr} \\
$m_{\Omega_c(\frac{1}{2}^-)}$&3.045~\cite{Yu:2022ymb}&$\langle g_{s}^{2}GG\rangle$&$0.47\pm0.15$ GeV$^4$~\cite{Narison:2010cg,Narison:2011xe,Narison:2011rn} \\
$m_e$&$0.511\times10^{-3}$~\cite{ParticleDataGroup:2024cfk}&$G_F$&$1.166\times10^{-5}$ GeV$^{-2}$~\cite{ParticleDataGroup:2024cfk} \\
$m_\mu$&$105.7\times10^{-3}$~\cite{ParticleDataGroup:2024cfk}&$V_{cb}$&0.041~\cite{ParticleDataGroup:2024cfk} \\
$m_\tau$&1.78~\cite{ParticleDataGroup:2024cfk}&~&~ \\			
\end{tabular}
\end{ruledtabular}
\end{table}

From the Eqs.~(\ref{eq:31}) and (\ref{eq:32}), one can find that the results of QCD sum rules for form factors depend on several parameters such as the Borel parameter $T^2$, continuum threshold parameters $s_0$ and $u_0$, and the square of transition momentum $Q^2$. The values of continuum threshold parameters are determined by the calculations of two-point QCDSR for single heavy baryon, which are $\sqrt{s_0}=6.60\pm0.10$, $6.70\pm0.10$ and $6.80\pm0.10$ GeV for $\Sigma_b$, $\Xi'_b$ and $\Omega_b$ and $\sqrt{u_0}=3.20\pm0.10$, $3.30\pm0.10$ and $3.40\pm0.10$ GeV for $\Sigma_c$, $\Xi'_c$ and $\Omega_c$~\cite{Wang:2009cr}. An appropriate work region of Borel parameter needs to be selected to obtain the final results. The form factors should have a weak Borel parameter dependency in this region. At the same time, the pole dominance and the convergence of OPE should be also satisfied. This work region is commonly called as `Borel platform'. To discuss the pole dominance, the following definitions for pole contributions of $s$ and $u$ channels are given as~\cite{Zhao:2020mod}:
\begin{eqnarray}\label{eq:35}
	\mathrm{Pole}_s=\frac{\int\limits_{u_{\min }}^{u_0} du \int\limits_{s_{\min }}^{s_0} ds}{\int\limits_{u_{\min }}^{u_0} du \int\limits_{s_{\min }}^{\infty} ds} ,
	\mathrm{Pole}_u=\frac{\int\limits_{u_{\min }}^{u_0} du \int\limits_{s_{\min }}^{s_0} ds}{\int\limits_{u_{\min }}^{\infty} du \int\limits_{s_{\min }}^{s_0} ds}.
\end{eqnarray}
The condition of pole dominance requires that the pole contributions for $s$ and $u$ channels should be both larger than 40\% in the present work.

\begin{figure}
	\centering
	\includegraphics[width=8.5cm]{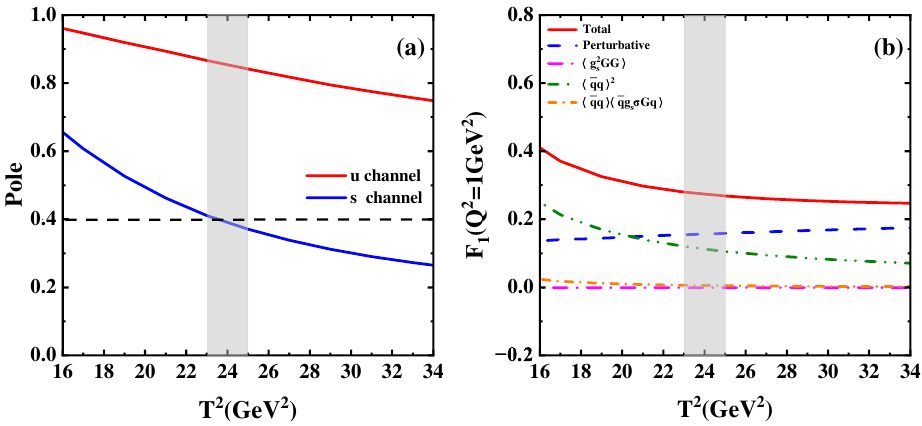}
	\caption{The pole contributions (a) and the contributions of perturbative term and different vacuum condensates (b) of form factor $F_1$ for $\Sigma_b\to\Sigma_c$ transition. The grey bound represent Borel platform.}
	\label{PCandOPEC}
\end{figure}

Taking the form factor $F_1$ for $\Sigma_b\to\Sigma_c$ transition as an example, we discuss how to determine the Borel platform. Fixing $Q^2=1$ GeV$^2$, we plot the pole contribution of form factors $F_1$ on Borel parameter $T^2$ (See Fig.~\ref{PCandOPEC} (a)). Besides, the contributions of perturbative part and different vacuum condensates for $F_1$ on Borel parameter $T^2$ are shown in Fig.~\ref{PCandOPEC} (b). For the $\Sigma_b\to\Sigma_c$ transition, the contributions from the vacuum condensate terms $\langle\bar{q}q\rangle$, $\langle\bar{q}g_s\sigma Gq\rangle$ and $\langle\bar{q}q\rangle\langle g_s^2GG\rangle$ can be neglected because they are proportional to the mass of $d$ quark. From the Fig.~\ref{PCandOPEC} (b), one can find that the contributions from $\langle g_s^2GG\rangle$ term are less than 1\%, The main contributions come from the perturbation part and the $\langle \bar{q}q \rangle^2$ term. In addition, the contribution from 8 dimension vacuum condensate term ($\langle\bar{q}q\rangle\langle\bar{q}g_s\sigma Gq\rangle$) is also tiny. The Borel platform of $F_1$ is determined as $23-25$ GeV$^2$. The trend of the form factor with the Borel parameter is relatively flat in this region. Besides, the pole contributions of $s$ and $u$ channels are about 40\% and 85\%, respectively and the contribution from $\langle\bar{q}q\rangle\langle\bar{q}g_s\sigma Gq\rangle$ term is about 2\%, which means the pole dominance and OPE convergence are both satisfied. After repeated trials, the Borel platforms for all form factors are determined and shown in Figs.~\ref{BWS}-\ref{BWO} in Appendix~\ref{Sec:AppA}. From these figures, it can be seen that on the corresponding Borel platforms, the dependence of the form factor $F_1$, $F_2$, $F_3$ and $G_3$ on the Borel parameter is weak. However, the $T^2$ dependence of form factors $G_1$ and $G_2$ is strong, which is due to the fact that the contributions from the perturbative part and the four quark condensate term have opposite signs, and the absolute values are close to each other. This leads $G_1$ and $G_2$ to be close to zero at $Q^2=1$ GeV$^2$. The impact of the numerical instability of $G_1$ and $G_2$ on the decay widths will be discussed in the following section.

\begin{figure}
	\centering
	\includegraphics[width=8.5cm]{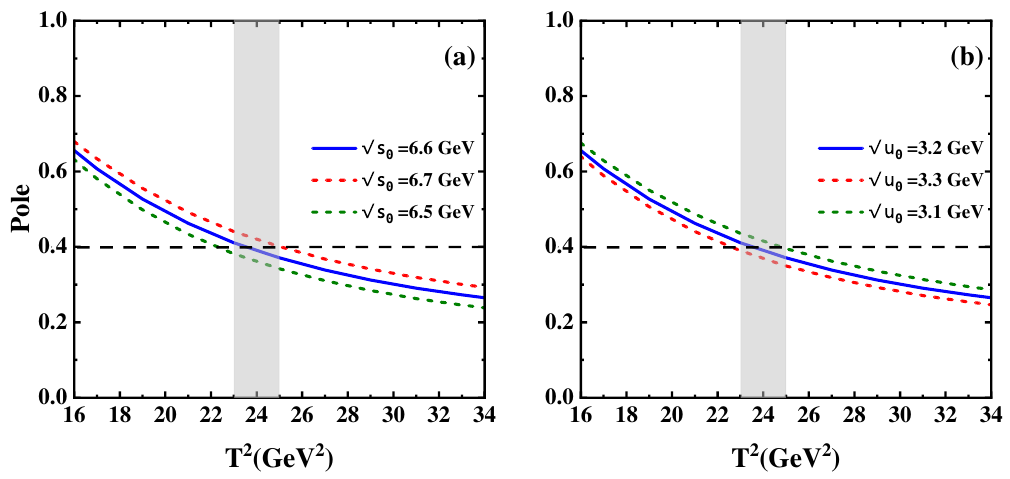}
	\caption{The $s$ channel pole contributions of the form factor $F_1$ for the $\Sigma_b \to \Sigma_c$ transition under different threshold parameters. In Fig.~(a), $\sqrt{u_0}=3.2$ GeV is fixed, while in Fig.~(b), $\sqrt{s_0}=6.6$ GeV is fixed. The grey bound indicates the Borel platform.}
	\label{PCR}
\end{figure}

The uncertainty analysis of the form factor is of great importance. According to Eqs. ~(\ref{eq:31}) and~(\ref{eq:32}), the uncertainties of the form factor originate from numerous input parameters, including baryon pole residues, Borel parameter, threshold parameters, vacuum condensate parameters and quark masses. It is worth noting that the threshold parameters and pole residues used in this work are obtained from previous two-point QCD sum rule calculations~\cite{Wang:2009cr}, and their uncertainties are closely correlated. If we consider their contributions simultaneously, the uncertainty in the form factor would be excessively amplified. Therefore, we only need to consider the uncertainty from the threshold parameters, as the uncertainty in the pole residues is already implicitly included. Furthermore, the choice of threshold parameters is closely related to the determination of the Borel parameters. Taking the form factor $F_1$ for $\Sigma_b\to\Sigma_c$ as an example, as shown in Fig.~\ref{PCandOPEC} (a), the pole contribution of $u$ channel is relatively large and varies slowly, while the Borel platform is mainly determined by the pole contribution of $s$ channel. We first plot the variation of the $s$ channel pole contribution with respect to the threshold parameters $s_0$ and $u_0$, as shown in Fig.~\ref{PCR}. From Fig.~\ref{PCR} (a), we observe that when $s_0$ takes its upper or lower bound, the Borel parameter should be adjusted accordingly so that the pole contribution satisfies $\mathrm{Pole}_s\approx40\%$. In other words, the upper and lower bounds of the Borel platform are used to determine the form factor values when $s_0$ is set to its upper and lower bounds, respectively. The determination of the uncertainty from $u_0$ follows a similar procedure. Since the contributions to the form factors mainly come from the perturbative term and the four quark condensate, the uncertainty arising from vacuum condensates is dominated by $\langle\bar{q}q\rangle$ ($\langle\bar{s}s\rangle$). Moreover, because the mass of the $s$ quark is much smaller than those of the $b$ and $c$ quarks, the uncertainty from quark masses is primarily governed by the $b$ and $c$ quarks. In summary, the uncertainties of the form factors mainly originate from the threshold parameters, quark condensate parameters and the masses of $b$ and $c$ quarks. The error estimates for all form factors at $Q^2=1$ GeV$^2$ are presented in Table~\ref{Error}. In this table, the central value of each form factors is listed first. For each input parameter, the upper deviation (superscript) is the change in the form factor when the parameter takes its upper bound, and the lower deviation (subscript) corresponds to the lower bound of the parameter.

\begin{table}[htbp]
	\begin{ruledtabular}\caption{Error estimates of the form factors at $Q^2 = 1$ GeV$^2$. The central value is firstly listed in the third column. For each input parameter, the upper deviation (superscript) is the change in the form factor when the parameter takes its upper bound, and the lower deviation (subscript) corresponds to the lower bound of the parameter.}
		\label{Error}
		\renewcommand\arraystretch{1.4}
		\begin{tabular}{c c c c c c c}
			Mode&FF&Central value&$\sqrt{s_0}$&$\sqrt{u_0}$&$\langle\bar{q}q\rangle(\langle\bar{s}s\rangle)$&$m_{b,c}$\\ \hline
			\multirow{6}*{$\Sigma_b \to \Sigma_c$}&$F_1$ &$0.27$&$~^{+0.04}_{-0.03}$&$~^{-0.02}_{+0.02}$&$~^{+0.01}_{-0.01}$&$~^{+0.01}_{-0.01}$ \\ 
			~&$F_2$&$0.35$&$~^{+0.04}_{-0.03}$&$~^{-0.02}_{+0.02}$&$~^{+0.01}_{-0.01}$&$~^{+0.01}_{-0.01}$ \\
			~&$F_3$&$-0.23$&$~^{+0.03}_{-0.03}$&$~^{-0.03}_{+0.03}$&$~^{+0.01}_{-0.01}$&$~^{+0.01}_{-0.01}$ \\
			~&$G_1$&$0.077$&$~^{+0.005}_{-0.004}$&$~^{+0.001}_{-0.001}$&$~^{+0.002}_{-0.002}$&$~^{+0.000}_{-0.000}$ \\
			~&$G_2$&$-0.075$&$~^{+0.004}_{-0.005}$&$~^{+0.002}_{-0.002}$&$~^{+0.003}_{-0.003}$&$~^{+0.001}_{-0.001}$ \\
			~&$G_3$&$-0.16$&$~^{+0.02}_{-0.03}$&$~^{-0.01}_{+0.01}$&$~^{+0.01}_{-0.01}$&$~^{+0.00}_{-0.00}$ \\ \hline
			\multirow{6}*{$\Xi'_b \to \Xi'_c$}&$F_1$ &$0.26$&$~^{+0.03}_{-0.03}$&$~^{-0.02}_{+0.02}$&$~^{+0.01}_{-0.01}$&$~^{+0.01}_{-0.01}$ \\
			~&$F_2$&$0.34$&$~^{+0.04}_{-0.03}$&$~^{-0.02}_{+0.01}$&$~^{+0.01}_{-0.01}$&$~^{+0.01}_{-0.01}$ \\
			~&$F_3$&$-0.21$&$~^{+0.03}_{-0.03}$&$~^{-0.03}_{+0.02}$&$~^{+0.01}_{-0.01}$&$~^{+0.01}_{-0.01}$ \\
			~&$G_1$&$0.068$&$~^{+0.003}_{-0.002}$&$~^{+0.003}_{-0.001}$&$~^{+0.002}_{-0.002}$&$~^{+0.000}_{-0.000}$ \\
			~&$G_2$&$-0.055$&$~^{+0.003}_{-0.003}$&$~^{+0.002}_{-0.002}$&$~^{+0.002}_{-0.002}$&$~^{+0.001}_{-0.001}$ \\
			~&$G_3$&$-0.16$&$~^{+0.02}_{-0.02}$&$~^{-0.00}_{+0.01}$&$~^{+0.01}_{-0.01}$&$~^{+0.00}_{-0.00}$ \\ \hline
			\multirow{6}*{$\Omega_b \to \Omega_c$}&$F_1$ &$0.24$&$~^{+0.03}_{-0.03}$&$~^{-0.02}_{+0.02}$&$~^{+0.01}_{-0.01}$&$~^{+0.01}_{-0.01}$ \\
			~&$F_2$&$0.31$&$~^{+0.04}_{-0.03}$&$~^{-0.02}_{+0.01}$&$~^{+0.01}_{-0.01}$&$~^{+0.01}_{-0.01}$ \\
			~&$F_3$&$-0.19$&$~^{+0.03}_{-0.03}$&$~^{-0.02}_{+0.03}$&$~^{+0.01}_{-0.01}$&$~^{+0.01}_{-0.01}$ \\
			~&$G_1$&$0.027$&$~^{+0.003}_{-0.002}$&$~^{+0.003}_{-0.001}$&$~^{+0.002}_{-0.002}$&$~^{+0.000}_{-0.000}$ \\
			~&$G_2$&$-0.0049$&$~^{+0.003}_{-0.002}$&$~^{+0.002}_{-0.002}$&$~^{+0.002}_{-0.001}$&$~^{+0.001}_{-0.000}$ \\
			~&$G_3$&$-0.15$&$~^{+0.02}_{-0.02}$&$~^{-0.00}_{+0.01}$&$~^{+0.01}_{-0.01}$&$~^{+0.00}_{-0.00}$ \\			
		\end{tabular}
	\end{ruledtabular}
\end{table}

\begin{figure*}
	\centering
	\includegraphics[width=18cm]{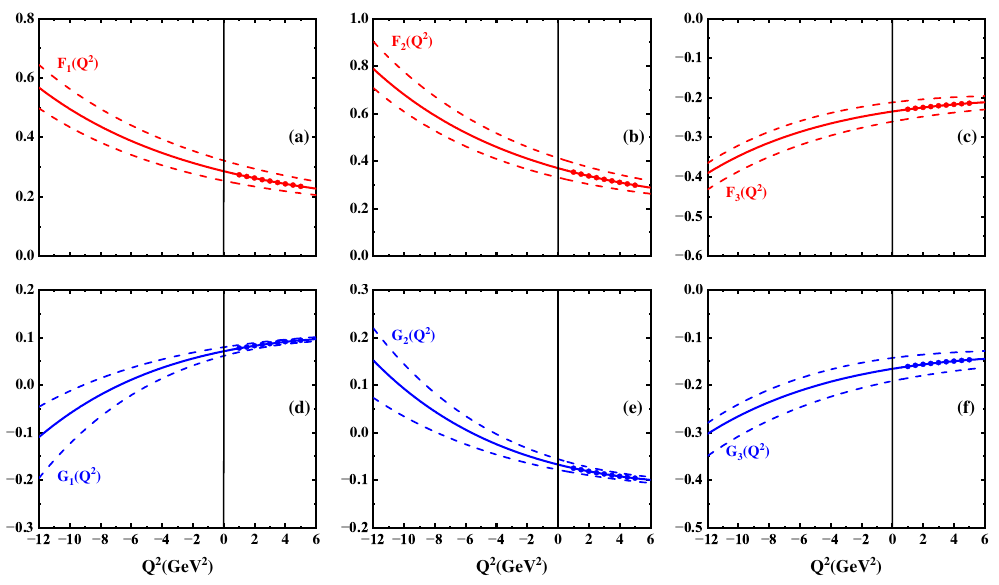}
	\caption{The fitting results of vector (a-c) and axial vector (d-f) form factors for $\Sigma_b\to\Sigma_c$ transition, where the red and blue real lines are fitting curves of central values for vector and axial vector form factors and the red and blue dash lines are fitting curves of upper and lower bounds.}
	\label{FFS}
\end{figure*}

\begin{figure*}
	\centering
	\includegraphics[width=18cm]{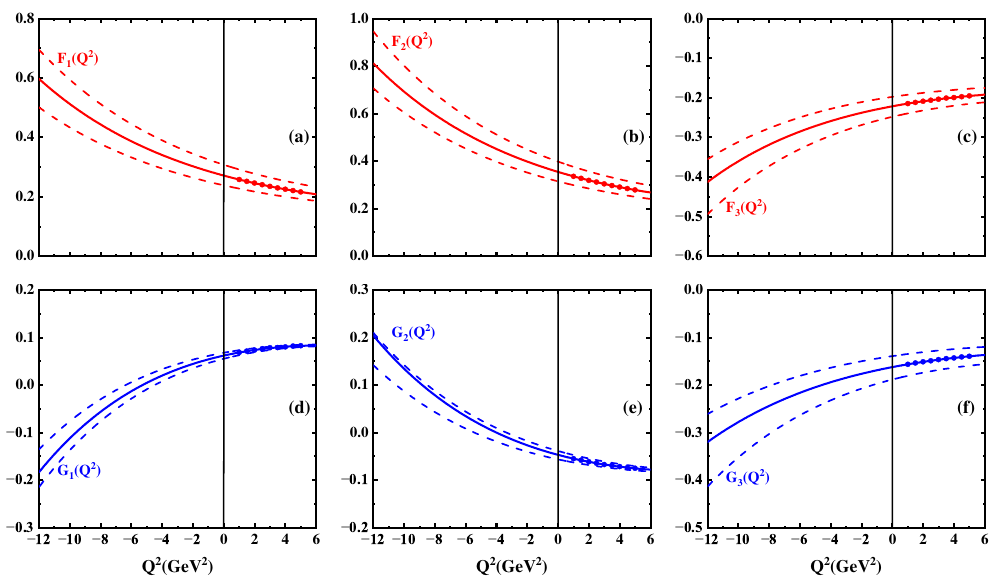}
	\caption{It is the same as Fig.~\ref{FFS}, but for $\Xi'_b\to\Xi'_c$ transition.}
	\label{FFXp}
\end{figure*}

\begin{figure*}
	\centering
	\includegraphics[width=18cm]{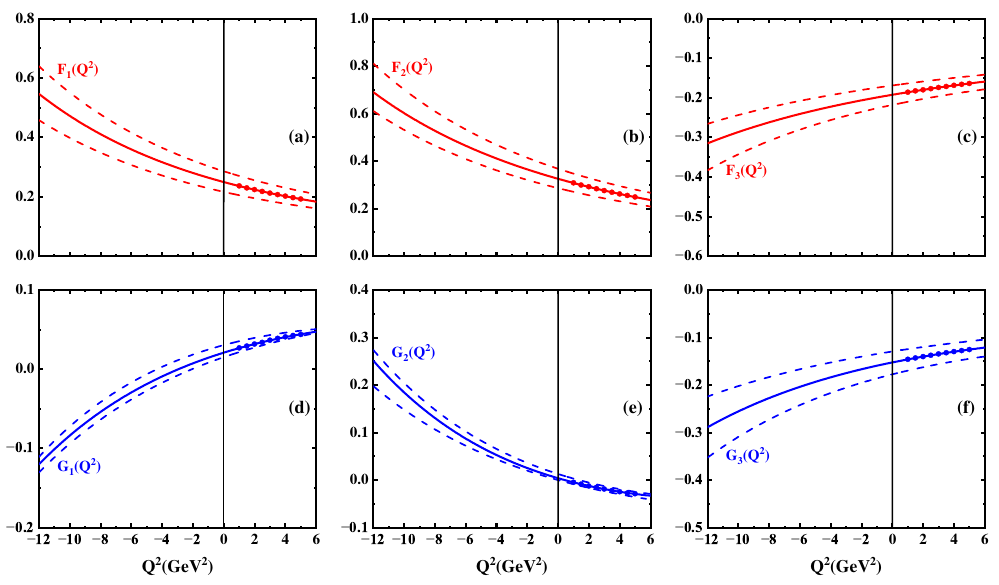}
	\caption{It is the same as Fig.~\ref{FFS}, but for $\Omega_b\to\Omega_c$ transition.}
	\label{FFO}
\end{figure*}

By taking the different values of $Q^2$, the form factors in space-like regions are obtained, where $Q^2$ is in the range of 1-5 GeV$^2$. The values of these form factors in time-like regions can be obtained by fitting the results in space-like regions with appropriate analytical function and extrapolating them into time-like regions. The $z$ series expansion approach is widely used to fit various form factors due to its high analytical, convergence, and low model dependence~\cite{Boyd:1994tt}. In the present work, the following $z$ series expansions for vector and axial vector form factors are employed, 
\begin{eqnarray}\label{eq:36}
\notag
F_i(Q^2) &&= \frac{F(0)}{1 + Q^2/m^2_{B_c^*(1^-)}}\\
\notag
&&\times \left\{ 1 + a\left[z(Q^2)-z(0)-\frac{1}{3}[z(Q^2)^3-z(0)^3] \right] \right.\\
&&\left.  + b\left[z(Q^2)^2-z(0)^2 + \frac{2}{3}[z(Q^2)^3-z(0)^3] \right] \right\},
\end{eqnarray}
\begin{eqnarray}\label{eq:37}
\notag
G_i(Q^2) &&= \frac{G(0)}{1 + Q^2/m^2_{B_c^*(1^+)}}\\
\notag
&&\times \left\{ 1 + a \left[z(Q^2)-z(0)-\frac{1}{3}[z(Q^2)^3 - z(0)^3] \right] \right.\\
&&\left.  + b \left[ z(Q^2)^2 - z(0)^2 + \frac{2}{3}[z(Q^2)^3 - z(0)^3] \right] \right\}.
\end{eqnarray}
Here the masses of ground state vector and axial vector $B_c$ meson are taken from the results of the modified Godfrey-Isgur quark model, which are $m_{B_c^*(1^-)}=6.338$ GeV and $m_{B_c^*(1^+)}=6.745$ GeV, respectively~\cite{Li:2023wgq}. $F(0)$, $G(0)$, $a$ and $b$ are fitting parameters, and function $z(Q^2)$ has the following form:
\begin{eqnarray}\label{eq:38}
z(Q^2) &&= \frac{\sqrt {t_+^2 + Q^2}  - \sqrt {t_+^2 - t_-^2}}{\sqrt {t_+^2 + Q^2}  + \sqrt {t_+^2 - t_-^2}},
\end{eqnarray}
where $t_{\pm}^2=(m_{\mathcal{B}_i}\pm m_{\mathcal{B}_f})^2$. Because the numerical results of the form factors exhibit a strong dependence on $Q^2$, including the errors at each $Q^2$ point in the fitting would render the results meaningless and would excessively amplify the uncertainties of the fit parameters. Therefore, in our analysis, we separately fitted the central value, the upper bound and the lower bound of the form factors for each point in the space-like region. This approach allows us to obtain the central value as well as the upper and lower bounds of the form factors in the physical region. The numerical results of these fitting parameters for the central values of all form factors are listed in Table~\ref{ZPC}. The fitting parameter tables of upper and lower bounds are shown in Appendix~\ref{Sec:AppB}. The fitting diagrams are explicitly shown in Figs.~\ref{FFS}-\ref{FFO}.
\begin{table}[htbp]
	\begin{ruledtabular}\caption{Fitting parameters for the central values of form factors (FF) using the $z$-series expansion, and the form factor values at $q^2= q_{\max}^2=(m_{\mathcal{B}_i}-m_{\mathcal{B}_f})^2$.}
		\label{ZPC}
		\renewcommand\arraystretch{1.3}
		\begin{tabular}{c c c c c c}
			Mode&FF&$F(0)$&$a$&$b$&$F(q^2_{\max})$\\ \hline
			\multirow{6}*{$\Sigma_b \to \Sigma_c$}&$F_1$ &$0.29$&$-10.45$&$53.62$&$0.55$ \\
			~&$F_2$ &$0.37$&$-13.26$&$69.66$&$0.75$ \\
			~&$F_3$ &$-0.23$&$-6.45$&$68.97$&$-0.37$\\
			~&$G_1$&$0.071$&$55.12$&$-249.51$&$-0.091$ \\
			~&$G_2$&$-0.067$&$68.66$&$-310.13$&$0.13$\\
			~&$G_3$&$-0.17$&$-10.92$&$87.72$&$-0.30$\\ \hline
			\multirow{6}*{$\Xi'_b \to \Xi'_c$}&$F_1$ &$0.27$&$-15.53$&$87.21$&$0.56$ \\
			~&$F_2$ &$0.35$&$-17.26$&$96.05$&$0.75$ \\
			~&$F_3$ &$-0.22$&$-10.61$&$95.33$&$-0.39$ \\
			~&$G_1$&$0.062$&$95.73$&$-629.57$&$-0.16$\\
			~&$G_2$&$-0.047$&$121.02$&$-694.22$&$0.18$\\
			~&$G_3$&$-0.16$&$-14.40$&$109.59$&$-0.30$\\ \hline
			\multirow{6}*{$\Omega_b \to \Omega_c$}&$F_1$ &$0.25$&$-15.02$&$65.72$&$0.52$ \\
			~&$F_2$ &$0.33$&$-12.89$&$39.32$&$0.66$ \\
			~&$F_3$ &$-0.19$&$-3.97$&$13.05$&$-0.30$ \\
			~&$G_1$&$0.021$&$145.24$&$-600.68$&$-0.11$\\
				~&$G_2$&$0.40\times10^{-2}$&$-1.35\times10^{3}$&$7.79\times10^{3}$&$0.22$\\
			~&$G_3$&$-0.15$&$-11.52$&$60.23$&$-0.27$
		\end{tabular}
	\end{ruledtabular}
\end{table}

The study of the form factors for $\Sigma_b\to\Sigma_c$ transition is relatively insufficient. However, the form factors for transitions $\Xi'_b\to\Xi'_c$ and $\Omega_b\to\Omega_c$ are analyzed by other collaboration in recent years. As a contrast, the values of these form factors at $Q^2=0$ from different collaborations are listed in Table~\ref{F0}. It can be found that the values of form factors $G_1$ and $G_2$ are close to zero in our prediction, and the form factors $F_{1,2,3}$ and $G_3$ for both $\Xi'_b\to\Xi'_c$ and $\Omega_b\to\Omega_c$ are close to each other, but slightly different. According to quark model, the single heavy baryons $\Sigma_{b(c)}$, $\Xi'_{b(c)}$ and $\Omega_{b(c)}$ have similar internal structure, the main difference lies in the numbers of $s$ quark it contains. Therefore, they should have same properties in light quarks SU(3) flavor symmetry. Our results clearly demonstrate this characteristic. The slightly difference of the values for form factors related $\Sigma_b\to\Sigma_c$, $\Xi'_b\to\Xi'_c$ and $\Omega_b\to\Omega_c$ mainly comes from the numbers and mass of $s$ quark. Since the mass of $s$ quark is much smaller than that of $b$ and $c$ quarks, the SU(3) flavor symmetry does not have obvious breaking. It is noted that the authors studied the semileptonic form factors for $\Xi'_b\to\Xi'_c$ and $\Omega_b\to\Omega_c$ by three-point QCDSR in Refs.~\cite{Neishabouri:2024gbc,Neishabouri:2025abl}. From Table~\ref{F0}, one can find that their results differ greatly from our predictions. Firstly, their results indicate that there is a significant difference in the values of the form factors between the $\Xi'_b\to\Xi'_c$ and $\Omega_b\to\Omega_c$ transitions, which indicates an obvious SU(3) symmetry breaking effect. Additionally, the authors show the effect of different Dirac structures on form factors in Ref.~\cite{Neishabouri:2024gbc} which are also listed in Table~\ref{F0}. It can be found that the different Dirac structures have a significant impact on the values of form factors. In the present work, all of the Dirac structures are considered and the form factors are determined by solving the twenty-four linear equations. Thus, the form factors are independence on Dirac structures and the results are more reliable and stable. Moreover, the semileptonic form factors for $\Omega_b\to\Omega_c$ are analyzed by LFQM in Ref.~\cite{Zhao:2018zcb}. The values of $f_{1,2}$ and $g_{1,2}$ (See Eq.~(\ref{eq:4})) at $Q^2=0$ are predicted as $f_1(0)=0.556$, $f_2(0)=0.531$, $g_1(0)=-0.170$ and $g_2(0)=-0.031$. By using the following relations between $f_i(g_i)$ and $F_i(G_i)$:
\begin{eqnarray}\label{eq:39}
	\notag
	f_1(q^2)&&=F_3(q^2)+\frac{m_{\mathcal{B}_i}+m_{\mathcal{B}_f}}{2m_{\mathcal{B}_i}}F_1(q^2)+\frac{m_{\mathcal{B}_i}+m_{\mathcal{B}_f}}{2m_{\mathcal{B}_f}}F_2(q^2),\\
	\notag
	f_2(q^2)&&=\frac{1}{2}\left[F_1(q^2)+\frac{m_{\mathcal{B}_i}}{m_{\mathcal{B}_f}}F_2(q^2)\right],\\
	\notag
	f_3(q^2)&&=\frac{1}{2}\left[F_1(q^2)-\frac{m_{\mathcal{B}_i}}{m_{\mathcal{B}_f}}F_2(q^2)\right],\\
	\notag
	g_1(q^2)&&=G_3(q^2)-\frac{m_{\mathcal{B}_i}-m_{\mathcal{B}_f}}{2m_{\mathcal{B}_i}}G_1(q^2)-\frac{m_{\mathcal{B}_i}-m_{\mathcal{B}_f}}{2m_{\mathcal{B}_f}}G_2(q^2),\\
	\notag
	g_2(q^2)&&=\frac{1}{2}\left[G_1(q^2)+\frac{m_{\mathcal{B}_i}}{m_{\mathcal{B}_f}}G_2(q^2)\right],\\
	g_3(q^2)&&=\frac{1}{2}\left[G_1(q^2)-\frac{m_{\mathcal{B}_i}}{m_{\mathcal{B}_f}}G_2(q^2)\right],
\end{eqnarray}
our predictions for the central values of form factors $f_{1,2}$ and $g_{1,2}$ at $Q^2=0$ are given as $f_1(0)=0.60$, $f_2(0)=0.58$, $g_1(0)=-0.14$ and $g_2(0)=-0.044$, which are comparable with the results of LFQM. 

\begin{table}[htbp]
	\begin{ruledtabular}\caption{The values of form factors (FF) at $Q^2=0$. The values in parentheses indicate the results obtained by selecting different Dirac structures.}
		\label{F0}
		\renewcommand\arraystretch{1.3}
		\begin{tabular}{c c c c c}
			Mode&FF&This work&Ref.~\cite{Neishabouri:2025abl}&Ref.~\cite{Neishabouri:2024gbc}\\ \hline
			\multirow{6}*{$\Xi'_b \to \Xi'_c$}&$F_1$ &$0.27^{+0.04}_{-0.03}$&$-0.45$&$-$ \\
			~&$F_2$ &$0.35^{+0.05}_{-0.04}$&$-0.18$&$-$ \\
			~&$F_3$ &$-0.22^{+0.02}_{-0.03}$&$0.44$&$-$ \\
			~&$G_1$ &$0.062^{+0.006}_{-0.007}$&$-0.75$&$$ \\
			~&$G_2$ &$-0.047^{+0.008}_{-0.009}$&$0.15$&$-$ \\
			~&$G_3$ &$-0.16^{+0.02}_{-0.03}$&$0.028$&$-$ \\ \hline
			\multirow{6}*{$\Omega_b \to \Omega_c$}&$F_1$ &$0.25^{+0.04}_{-0.03}$&$-$&$0.39(-0.13,0.66)$ \\
			~&$F_2$ &$0.33^{+0.04}_{-0.04}$&$-$&$-0.04(0.57,0.37)$ \\
			~&$F_3$ &$-0.19^{+0.02}_{-0.03}$&$-$&$-0.28$ \\
			~&$G_1$ &$0.021^{+0.009}_{-0.006}$&$-$&$0.40$ \\
			~&$G_2$ &$(0.40^{+0.89}_{-0.40})\times10^{-2}$&$-$&$-0.037(0.44,-0.60)$ \\
			~&$G_3$ &$-0.15^{+0.02}_{-0.03}$&$-$&$-0.018$ \\
		\end{tabular}
	\end{ruledtabular}
\end{table}

\begin{figure*}
	\centering
	\includegraphics[width=18cm]{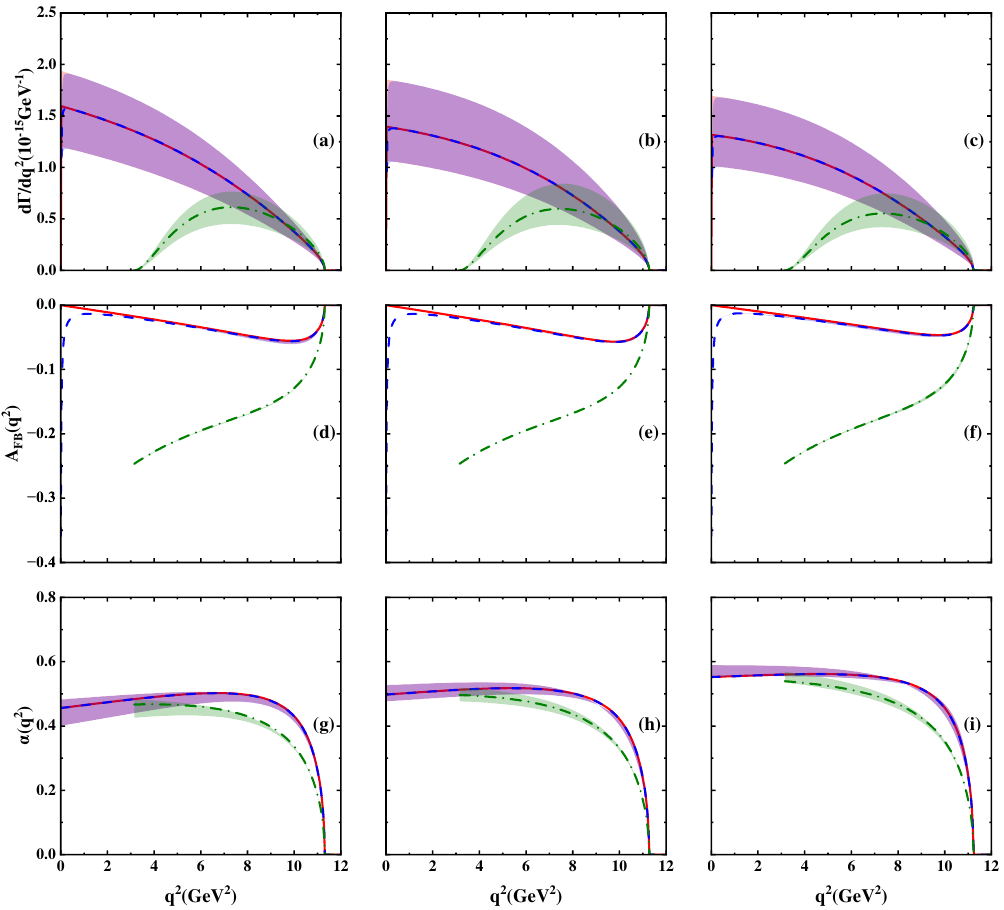}
	\caption{The differential decay width $d\Gamma/dq^{2}$, leptonic forward-backward asymmetry $A_{FB}$ and asymmetry parameter $\alpha$ with variations of $q^{2}$ for semileptonic decay processes $\Sigma_{b}\to \Sigma_{c}l\bar{\nu}_{l}$ (a, d and g), $\Xi'_{b}\to\Xi'_{c}l\bar{\nu}_{l}$ (b, e and h) and $\Omega_{b}\to\Omega_{c}l\bar{\nu}_{l}$ (c, f and i), where the red, blue and green represent $l=e$, $\mu$ and $\tau$, respectively.}
	\label{SEMI}
\end{figure*}

With the above form factors, the semileptonic decay processes $\Sigma_b\to\Sigma_cl\bar{\nu}_l$, $\Xi'_b\to\Xi'_cl\bar{\nu}_l$ and $\Omega_b\to\Omega_cl\bar{\nu}_l$ can be analyzed. Firstly, according to the Eqs.~(\ref{eq:5})-(\ref{eq:8}) and (\ref{eq:39}), the differential decay widths with variations of $q^2$ can be obtained and are shown in Figs.~\ref{SEMI} (a)-(c). The numerical results for decay widths are collected in Table~\ref{Width}, where the errors in the decay widths mainly origin from the form factors. Besides, the values of longitudinally, transversely polarized decay widths and their ratio are also shown in Table~\ref{Width}. According to SU(3) flavor symmetry, the total decay widths of $\Sigma_b\to\Sigma_cl\bar{\nu}_l$, $\Xi'_b\to\Xi'_cl\bar{\nu}_l$ and $\Omega_b\to\Omega_cl\bar{\nu}_l$ should satisfy the relation:
\begin{eqnarray}\label{eq:40}
	\notag
	\Gamma(\Sigma^-_b\to\Sigma^0_cl^-\bar{\nu}_l)=\Gamma(\Xi'^-_b\to\Xi'^0_cl^-\bar{\nu}_l)=\Gamma(\Omega^-_b\to\Omega^0_cl^-\bar{\nu}_l).\\
\end{eqnarray}
We have compared the predictions of our QCD sum rule results with those of SU(3) flavor symmetry relation, which can be seen in Table~\ref{WidthSU3}. Our sum rule results indicate that the SU(3) flavor symmetry is slightly broken in these decay processes, and can reach about 1$\sim$10 \%. Dramatically, the QCDSR results of other collaboration indicate that $\Gamma(\Omega^-_b\to\Omega^0_c(e^-,\mu^-)\bar{\nu}_{e,\mu})\approx2\Gamma(\Xi'^-_b\to\Xi'^0_c(e^-,\mu^-)\bar{\nu}_{e,\mu})$ and $\Gamma(\Omega^-_b\to\Omega^0_c\tau^-\bar{\nu}_\tau)\approx5\Gamma(\Xi'^-_b\to\Xi'^0_c\tau^-\bar{\nu}_\tau)$, the SU(3) symmetry is severely broken~\cite{Neishabouri:2024gbc,Neishabouri:2025abl}. These differences require further experimental and theoretical verification.
\begin{table}[htbp]
	\begin{ruledtabular}\caption{The total, longitudinally and transversely polarized decay widths ($\Gamma$, $\Gamma_L$ and $\Gamma_T$) in $10^{-15}$ GeV and the ratio of longitudinally and transversely polarized decay widths $\Gamma_L/\Gamma_T$ for semileptonic decays $\Sigma^-_b\to\Sigma^0_cl^-\bar{\nu}_l$, $\Xi'^-_b\to\Xi'^0_cl^-\bar{\nu}_l$ and $\Omega^-_b\to\Omega^0_cl^-\bar{\nu}_l$. }
		\label{Width}
		\renewcommand\arraystretch{1.3}
		\begin{tabular}{c c c c c c}
			Decay channel&$\Gamma$&$\Gamma_L$&$\Gamma_T$&$\Gamma_L/\Gamma_T$&$\Gamma$~\cite{Neishabouri:2024gbc,Neishabouri:2025abl} \\ \hline
			$\Sigma_b^-\to\Sigma_c^0e^-\bar{\nu}_e$&$11.29^{+2.65}_{-2.92}$&$10.04^{+2.35}_{-2.58}$&$1.25^{+0.29}_{-0.34}$&$8.03^{+2.88}_{-2.78}$&$-$ \\
			$\Sigma_b^-\to\Sigma_c^0\mu^-\bar{\nu}_\mu$&$11.25^{+2.64}_{-2.91}$&$10.00^{+2.35}_{-2.57}$&$1.25^{+0.29}_{-0.34}$&$8.00^{+2.88}_{-2.77}$&$-$\\
			$\Sigma_b^-\to\Sigma_c^0\tau^-\bar{\nu}_\tau$&$3.39^{+0.83}_{-0.89}$&$2.98^{+0.74}_{-0.78}$&$0.41^{+0.09}_{-0.11}$&$7.27^{+2.66}_{-2.48}$&$-$ \\ \hline
			$\Xi'^-_b\to\Xi'^0_ce^-\bar{\nu}_e$&$10.53^{+4.14}_{-2.65}$&$9.28^{+3.43}_{-2.31}$&$1.25^{+0.71}_{-0.34}$&$7.42^{+3.41}_{-4.60}$&$5.01$ \\
			$\Xi'^-_b\to\Xi'^0_c\mu^-\bar{\nu}_\mu$&$10.49^{+4.13}_{-2.64}$&$9.25^{+3.42}_{-2.31}$&$1.24^{+0.71}_{-0.33}$&$7.46^{+3.40}_{-4.67}$&$4.92$ \\
			$\Xi'^-_b\to\Xi'^0_c\tau^-\bar{\nu}_\tau$&$3.30^{+1.37}_{-0.87}$&$2.88^{+1.12}_{-0.75}$&$0.42^{+0.25}_{-0.12}$&$6.86^{+3.31}_{-4.46}$&$0.61$ \\ \hline
			$\Omega^-_b\to\Omega^0_ce^-\bar{\nu}_e$&$9.78^{+3.34}_{-2.34}$&$8.70^{+2.80}_{-2.01}$&$1.08^{+0.54}_{-0.33}$&$8.06^{+3.58}_{-4.44}$&$11.00$ \\
			$\Omega^-_b\to\Omega^0_c\mu^-\bar{\nu}_\mu$&$9.75^{+3.33}_{-2.34}$&$8.67^{+2.79}_{-2.00}$&$1.08^{+0.53}_{-0.33}$&$8.03^{+3.56}_{-4.35}$&$11.00$ \\
			$\Omega^-_b\to\Omega^0_c\tau^-\bar{\nu}_\tau$&$3.03^{+1.10}_{-0.74}$&$2.67^{+0.92}_{-0.62}$&$0.36^{+0.19}_{-0.12}$&$7.42^{+3.56}_{-4.28}$&$3.11$ \\
		\end{tabular}
	\end{ruledtabular}
\end{table}

\begin{table}[htbp]
	\begin{ruledtabular}\caption{Quantitative predictions of SU(3) flavor symmetry breaking for semileptonic decays of single bottom baryons, where the decay widths are in unit $10^{-15}$ GeV. }
		\label{WidthSU3}
		\renewcommand\arraystretch{1.3}
		\begin{tabular}{c c c c}
			Decay channel&$\Gamma$ (QCDSR)&$\Gamma$ (SU(3))&$\left|\frac{\mathrm{QCDSR}-\mathrm{SU(3)}}{\mathrm{SU(3)}}\right|$ \\ \hline
			$\Sigma_b^-\to\Sigma_c^0e^-\bar{\nu}_e$&$11.29$&$11.29$&$-$ \\
			$\Sigma_b^-\to\Sigma_c^0\mu^-\bar{\nu}_\mu$&$11.25$&$11.25$&$-$ \\
			$\Sigma_b^-\to\Sigma_c^0\tau^-\bar{\nu}_\tau$&$3.39$&$3.39$&$-$ \\ \hline
			$\Xi'^-_b\to\Xi'^0_ce^-\bar{\nu}_e$&$10.53$&$11.29$&$6.73\%$ \\
			$\Xi'^-_b\to\Xi'^0_c\mu^-\bar{\nu}_\mu$&$10.49$&$11.25$&$6.76\%$ \\
			$\Xi'^-_b\to\Xi'^0_c\tau^-\bar{\nu}_\tau$&$3.30$&$3.39$&$2.65\%$ \\ \hline
			$\Omega^-_b\to\Omega^0_ce^-\bar{\nu}_e$&$9.78$&$11.29$&$13.37\%$ \\
			$\Omega^-_b\to\Omega^0_c\mu^-\bar{\nu}_\mu$&$9.75$&$11.25$&$13.33\%$ \\
			$\Omega^-_b\to\Omega^0_c\tau^-\bar{\nu}_\tau$&$3.03$&$3.39$&$10.62\%$ \\
		\end{tabular}
	\end{ruledtabular}
\end{table}

In our view, introducing contributions from negative parity baryons in the correlation functions on the phenomenological side is the only correct approach to extract hadronic parameters using three-point QCD sum rules, as it eliminates the dependence of the results on the Dirac structure. In previous sum rule studies~\cite{Neishabouri:2024gbc,Neishabouri:2025abl}, the authors did not consider such contributions, which led to a strong dependence of their form factors on the chosen Dirac structures. In other words, without including the contributions from negative parity hadronic states, the form factors extracted on the hadronic side can be obtained by selecting different combinations of Dirac structures on the QCD side. For a given form factor, different Dirac-structure combinations may yield completely different predictions, rendering the final results meaningless(See Table~\ref{F0}). It is well known that the $\Xi'_{b(c)}$ and $\Omega_{b(c)}$ baryons have similar internal structures, differing mainly in the numbers of $s$ quark they contain. According to SU(3) flavor symmetry, the semileptonic decay widths of $\Xi'_b\to\Xi'_cl\bar{\nu}_l$ and $\Omega_b\to\Omega_cl\bar{\nu}_l$ should not differ significantly. However, the widths predicted in Refs.~\cite{Neishabouri:2024gbc,Neishabouri:2025abl} differ drastically, which is inconsistent with the intuitive physical picture. Moreover, the calculation method used on the QCD side in Refs.~\cite{Neishabouri:2024gbc,Neishabouri:2025abl} is completely different from ours. We employ the Cutkosky's rules to obtain the spectral density on the QCD side, which is a method widely adopted in three-point sum rule calculations. It is also worth noting that early influential works using QCD sum rules and light-cone QCD sum rules to study baryon masses and form factors such as  already considered the crucial point of including contributions from opposite parity baryons~\cite{Khodjamirian:2011jp,Jido:1996ia}. In summary, there are multiple reasons for the significant discrepancy between our results and previous QCD sum rule predictions regarding the extent of SU(3) flavor symmetry breaking. It is certain that the inclusion of negative parity baryon contributions is necessary.

\begin{figure}
	\centering
	\includegraphics[width=8.5cm]{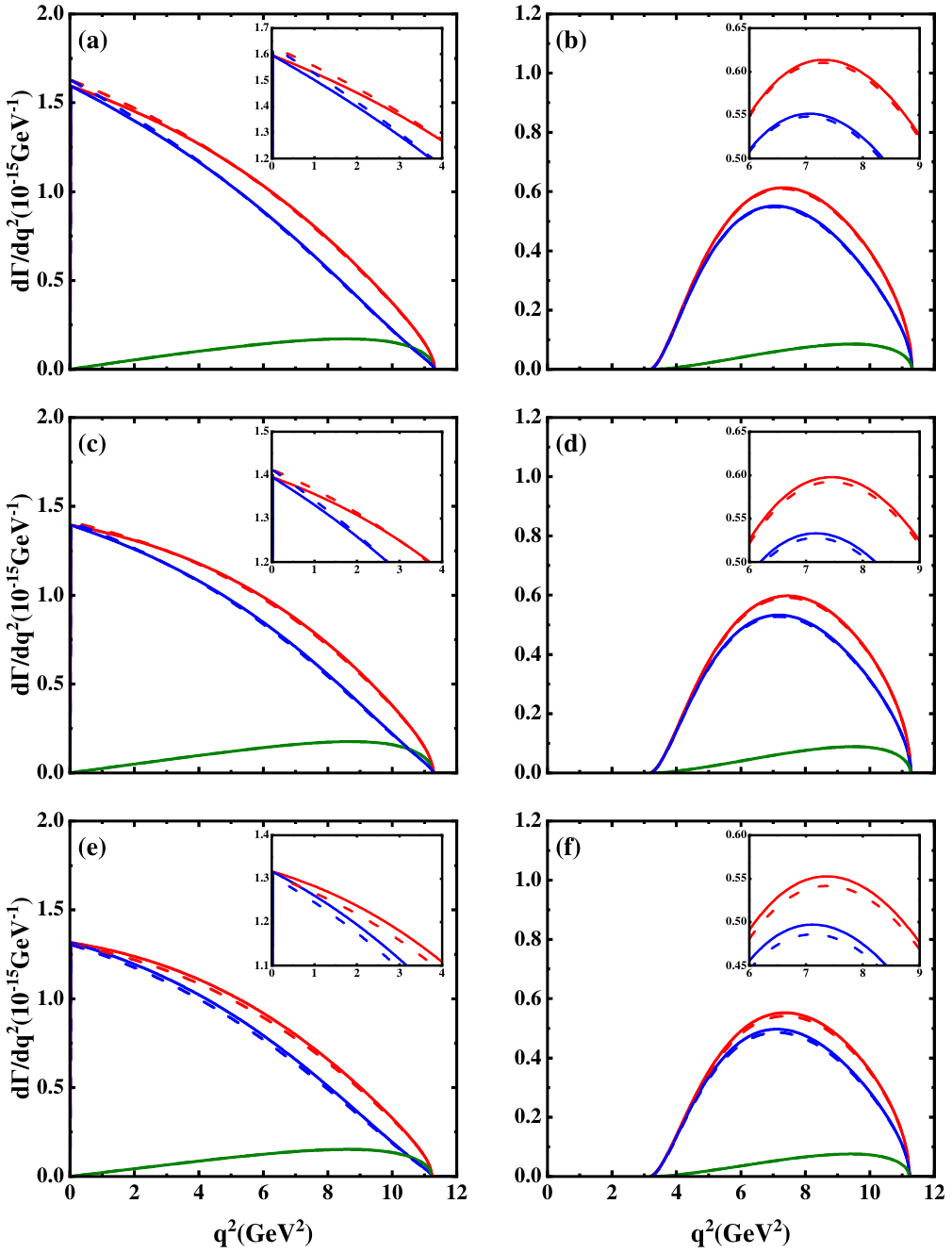}
	\caption{The total (red), longitudinal (blue), and transverse (green) differential decay widths with variations of $q^2$ for semileptonic decays $\Sigma_b\to\Sigma_cl\bar{\nu}_l$ (a, b), $\Xi'_b\to\Xi'_cl\bar{\nu}_l$ (c, d) and $\Omega_b\to\Omega_cl\bar{\nu}_l$ (e, f), where the (a), (c) and (e) are electron channel and (b), (d) and (f) are $\tau$ channel. The dash lines denote the results when $G_1=G_2=0$.}
	\label{DWR}
\end{figure}

As mentioned above, the numerical values of the form factors $G_1$ and $G_2$ at $Q^2=1$
GeV$^2$ exhibit strong dependence on the Borel parameter, which originates from the cancellation between the perturbative contribution and the four quark condensate. It is observed that these two form factors are close to zero in the space-like region. However, their fitted results indicate that their contributions in the physical time-like region can not be neglected. To analyze their impact on the decay width, we take the limit 
$G_{1,2}\to0$ and plot the corresponding total, longitudinal, and transverse differential decay widths as functions of $q^2$, comparing them with the original results, as shown in Fig.~\ref{DWR}. From Fig.~\ref{DWR}, it can be seen that for semileptonic decays $\Sigma_b\to\Sigma_cl\bar{\nu}_l$, $\Xi'_b\to\Xi'_cl\bar{\nu}_l$ and $\Omega_b\to\Omega_cl\bar{\nu}_l$, when $G_1$ and $G_2$ are set to zero, the transverse differential decay width remains practically unaffected, and the change in the longitudinal differential decay width is very small. Hence, the total differential width experiences a similarly minimal variation. 

The PDG reports that $\Sigma^-_b$ and $\Xi'^-_b$ baryons have relatively large total decay widths of $5.3\pm0.5$ and $0.03\pm0.032$ MeV, respectively~\cite{ParticleDataGroup:2024cfk}. This is because they have strong decay channels such as $\Sigma^-_b\to\Lambda^0_b\pi^-$ and $\Xi'^-_b\to\Xi^0_b\pi^-$. According to our results, the branching ratios of semileptonic decays $\Sigma_b\to\Sigma_cl\bar{\nu}_l$ and $\Xi'_b\to\Xi'_cl\bar{\nu}_l$ are in the order of magnitude $10^{-12}$ and $10^{-10}$, which seems difficult to be measured by current experiments. However, the $\Omega_b$ baryon has a lifetime close to that of the $\Lambda_b$ and $\Xi_b$ baryons, thus the study of its weak decay processes is of great significance. In recent years, some research groups have used different theoretical models to study the weak decay channels of the $\Omega_b$ baryon. For comparison, our predictions for the values of the $\Omega^-_b$ semileptonic decay branching ratios and these of other collaboration's are collected in Table~\ref{Br}. From Table~\ref{Br}, one can find that the central value we obtained for $\mathcal{B}(\Omega^-_b\to\Omega^0_ce^-\bar{\nu}_e)$ is close to the results of theoretical models including QCDSR~\cite{Neishabouri:2024gbc}, the relativistic quark model (RQM)~\cite{Ivanov:1999pz}, LFQM~\cite{Zhao:2018zcb} and Bethe-Salpeter equation (BSE)~\cite{Ivanov:1998ya}, but smaller than the results from the heavy quark effective theory (HQET)~\cite{Xu:1992hj}, Large $N_c$ expand approach (Large $N_c$)~\cite{Du:2011nj} and non-relativistic quark model (NRQM)~\cite{Cheng:1995fe}. For the $\tau$ channel, our prediction for branching ratio is close to the results from other QCDSR studies, but respectively larger and smaller than the results from HQET~\cite{Xu:1992hj} and Large $N_c$~\cite{Han:2020sag}. 

\begin{table}[htbp]
	\begin{ruledtabular}\caption{The branching ratios (BR) (in \%) of semileptonic decays $\Omega^-_b\to\Omega^0_cl^-\bar{\nu}_l$, which are calculated at $\tau_{\Omega^-_b}=(1.64\pm0.16)\times10^{-12}$ s~\cite{ParticleDataGroup:2024cfk}.}
		\label{Br}
		\renewcommand\arraystretch{1.3}
		\begin{tabular}{c c c }
			BR&$\mathcal{B}(\Omega^-_b\to\Omega^0_ce^-\bar{\nu}_e)$&$\mathcal{B}(\Omega^-_b\to\Omega^0_c\tau^-\bar{\nu}_\tau)$ \\ \hline
			This work&$2.44^{+0.83}_{-0.59}$&$0.75^{+0.28}_{-0.18}$\\
			QCDSR~\cite{Neishabouri:2024gbc}&$2.74$&$0.78$\\
			HQET~\cite{Xu:1992hj}&$3.29$&$0.52$\\
			Large $N_c$~\cite{Du:2011nj,Han:2020sag}&$4.20$&$1.20$\\
			BSE~\cite{Ivanov:1998ya}&$2.97$&$-$\\
			NRQM~\cite{Cheng:1995fe}&$3.76$&$-$\\
			RQM~\cite{Ivanov:1999pz}&$2.77$&$-$\\
			LFQM~\cite{Zhao:2018zcb}&$2.84$&$-$\\
		\end{tabular}
	\end{ruledtabular}
\end{table}

Furthermore, we also analyze some parameters related to new physics in these decay processes, including the lepton universality ratio $R_{\mathcal{B}_f}$, leptonic forward-backward asymmetry $A_{FB}$ and asymmetry parameter $\alpha$. The definitions of these parameters are as follows:
\begin{eqnarray}\label{eq:41}
	R_{\mathcal{B}_f}&&=\frac{\Gamma(\mathcal{B}_i\to\mathcal{B}_f\tau\bar{\nu}_\tau)}{\Gamma(\mathcal{B}_i\to\mathcal{B}_f e\bar{\nu}_e)},
\end{eqnarray}
\begin{eqnarray}\label{eq:42}
	\notag
	A_{FB}(q^2)&&=\frac{d\Gamma_{\mathrm{forward}}/dq^2-d\Gamma_{\mathrm{backward}}/dq^2}{d\Gamma/dq^2}\\
	\notag
	&&=\frac{3}{4}\left({\frac{|H_{\frac{1}{2},1}|^2-|H_{-\frac{1}{2},-1}|^2}{|H_{\mathrm{tot}}|^2}}\right.\\
	&&\left.{-\frac{2m_l^2}{q^2}\frac{H_{\frac{1}{2},0}H_{\frac{1}{2},t}^*+H_{-\frac{1}{2},0}H_{-\frac{1}{2},t}^*}{|H_{\mathrm{tot}}|^2}}\right),
\end{eqnarray}
\begin{eqnarray}\label{eq:43}
	\alpha&&=\frac{d\Gamma_{\lambda_2=\frac{1}{2}}/dq^2-d\Gamma_{\lambda_2=-\frac{1}{2}}/dq^2}{d\Gamma_{\lambda_2=\frac{1}{2}}/dq^2+d\Gamma_{\lambda_2=-\frac{1}{2}}/dq^2},
\end{eqnarray}
where,
\begin{eqnarray}\label{eq:44}
	\notag
	|H_{\mathrm{tot}}|^2&&=\left(1+\frac{m_l^2}{2q^2}\right)\left(|H_{\frac{1}{2},1}|^2+|H_{-\frac{1}{2},-1}|^2{+|H_{\frac{1}{2},0}|^2}\right.\\
	\notag
	&&\left.{+|H_{-\frac{1}{2},0}|^2}\right)+\frac{3m_l^2}{2q^2}\left(|H_{\frac{1}{2},t}|^2+|H_{-\frac{1}{2},t}|^2\right),\\
	\notag
	\frac{d\Gamma _{\lambda_2=\pm\frac{1}{2}}}{dq^2} &&= \frac{G_F^2V_{cb}^2q^2}{384\pi ^3m_{\mathcal{B}_i}^2}\frac{\sqrt{Q_ + Q_ - }}{2m_{\mathcal{B}_i}}\left(1 - \frac{m_l^2}{q^2} \right)^2\\
	\notag
	&&\times \left[ \frac{4m_l^2}{3q^2}\left(|H_{\pm\frac{1}{2},1}|^2 + |H_{\pm\frac{1}{2},0}|^2+3|H_{\pm\frac{1}{2},t}|^2\right) \right.\\
	&&\left. { + \frac{8}{3}\left(|H_{\pm\frac{1}{2},1}|^2 + |H_{\pm\frac{1}{2},0}|^2\right)} \right].
\end{eqnarray}
The lepton universality ratio $R_{\mathcal{B}_f}$ of these decay processes are shown in the last column of Table~\ref{Decay}. From Table~\ref{Decay}, one can find that the central values of our results for these decay processes are all close to $0.30$. The $q^2$ dependence of of $A_{FB}$ and $\alpha$ for these decay processes are shown in Figs.~\ref{SEMI} (d)-(i). From Figs.~\ref{SEMI} (d)-(f), we can see that the leptonic forward-backward asymmetry $A_{FB}$ of these decay modes all becomes $0$ near the zero recoil point $q^2=(m_{\mathcal{B}_f}-m_{\mathcal{B}_i})^2$ for different leptonic decay channels. However, the behaviors of different leptonic decay channels are clearly different near the maximum recoil point $q^2=m_l^2$. The central value of $A_{FB}$ is going to $0$ and $-0.24$ for the electric and $\mu(\tau)$ decay channels, respectively. For the asymmetric parameter $\alpha$, its central value increases from $0.46$ to $0$, $0.5$ to $0$ and $0.55$ to $0$ as $q^2$ increases from $m_l^2$ to $(m_{\mathcal{B}_f}-m_{\mathcal{B}_i})^2$ for the decay processes $\Sigma_b\to\Sigma_cl\bar{\nu}_l$, $\Xi'_b\to\Xi'_cl\bar{\nu}_l$ and $\Omega_b\to\Omega_cl\bar{\nu}_l$, respectively (See Figs.~\ref{SEMI} (g)-(i)).
After integrating the numerators and denominators in Eqs.~(\ref{eq:42}) and~(\ref{eq:43}) separately over the physical $q^2$ region, the mean values of $A_{FB}$ and $\alpha$ are obtained. The numerical results for these asymmetry parameters are collected in Table~\ref{Decay}. Future high energy physics experiments can measure these parameters and compare them with our predictions, which can help deepen understand of the decay properties for single heavy baryons. Moreover, the measurement of these parameters can also provide the possibility to explore new physics beyond the SM.

\begin{table}[htbp]
	\begin{ruledtabular}\caption{The mean values of leptonic forward-backward asymmetry $\langle A_{FB} \rangle$ and asymmetry parameter $\langle\alpha\rangle$ and the lepton universality ratios $R_{\mathcal{B}_f}$ for semileptonic decays $\Sigma^-_b\to\Sigma^0_cl^-\bar{\nu}_l$, $\Xi'^-_b\to\Xi'^0_cl^-\bar{\nu}_l$ and $\Omega^-_b\to\Omega^0_cl^-\bar{\nu}_l$. }
		\label{Decay}
		\renewcommand\arraystretch{1.3}
		\begin{tabular}{c c c c}
			Decay channel&$\langle A_{FB}\rangle$&$\langle \alpha\rangle$&$R_{\mathcal{B}_f}$ \\ \hline
			$\Sigma_b^-\to\Sigma_c^0e^-\bar{\nu}_e$&$-0.025^{+0.000}_{-0.001}$&$0.48^{+0.01}_{-0.04}$&\multirow{3}*{$0.30^{+0.11}_{-0.11}$} \\
			$\Sigma_b^-\to\Sigma_c^0\mu^-\bar{\nu}_\mu$&$-0.030^{+0.000}_{-0.001}$&$0.48^{+0.01}_{-0.04}$&~ \\
			$\Sigma_b^-\to\Sigma_c^0\tau^-\bar{\nu}_\tau$&$-0.17^{+0.00}_{-0.00}$&$0.42^{+0.00}_{-0.02}$&~ \\ \hline
			$\Xi'^-_b\to\Xi'^0_ce^-\bar{\nu}_e$&$-0.025^{+0.000}_{-0.002}$&$0.50^{+0.02}_{-0.02}$&\multirow{3}*{$0.31^{+0.15}_{-0.15}$} \\
			$\Xi'^-_b\to\Xi'^0_c\mu^-\bar{\nu}_\mu$&$-0.031^{+0.000}_{-0.001}$&$0.50^{+0.02}_{-0.02}$&~ \\
			$\Xi'^-_b\to\Xi'^0_c\tau^-\bar{\nu}_\tau$&$-0.17^{+0.00}_{-0.00}$&$0.43^{+0.01}_{-0.02}$&~ \\ \hline
			$\Omega^-_b\to\Omega^0_ce^-\bar{\nu}_e$&$-0.021^{+0.000}_{-0.002}$&$0.55^{+0.02}_{-0.01}$&\multirow{3}*{$0.31^{+0.13}_{-0.13}$} \\
			$\Omega^-_b\to\Omega^0_c\mu^-\bar{\nu}_\mu$&$-0.027^{+0.001}_{-0.002}$&$0.55^{+0.01}_{-0.01}$&$$ \\
			$\Omega^-_b\to\Omega^0_c\tau^-\bar{\nu}_\tau$&$-0.17^{+0.00}_{-0.00}$&$0.46^{+0.01}_{-0.01}$&$$ \\
		\end{tabular}
	\end{ruledtabular}
\end{table}

\section{Conclusion}\label{sec5}
In this work, we firstly investigate the electroweak transition form factors of baryon processes $\Sigma_b\to\Sigma_c$, $\Xi'_b\to\Xi'_c$ and $\Omega_b\to \Omega_c$ within the framework of three-point QCDSR. In phenomenological side, all possible couplings of interpolating current to hadronic states are considered. And the Dirac structure dependence of the form factors is systematically eliminated. In QCD side, the OPE is truncated at dimension of 8, thereby including contributions from a wider set of Feynman diagrams and improving the reliability of the calculation. With the estimated form factors, we study the semileptonic decay processes $\Sigma_b\to\Sigma_cl\bar{\nu}_l$, $\Xi'_b\to\Xi'_cl\bar{\nu}_l$ and $\Omega_b\to \Omega_cl\bar{\nu}_l$. Our results indicate a slight breaking of SU(3) flavor symmetry in these processes. Moreover, the lepton universality ratio $R_{\mathcal{B}_f}$, the leptonic forward-backward asymmetry $A_{FB}$ and the asymmetry parameter $\alpha$ are also given, which can offer valuable probes for potential new physics beyond the SM. Finally, we expect those results will serve as a useful reference for future theoretical and experimental studies of weak decays involving heavy flavor baryons.

\section*{Acknowledgments}
This work is supported by National Natural Science Foundation of China under the Grant No. 12575083, as well as supported, in part, by National Key Research and Development Program under Grant No. 2024 YFA1610503 and Natural Science Foundation of HeBei Province under the Grant No. A2024502002.

\appendix
\section{The contributions of perturbative part and different vacuum condensate terms for all form factors in corresponding Borel platforms.}\label{Sec:AppA}
\begin{figure*}
		\centering
		\includegraphics[width=18cm]{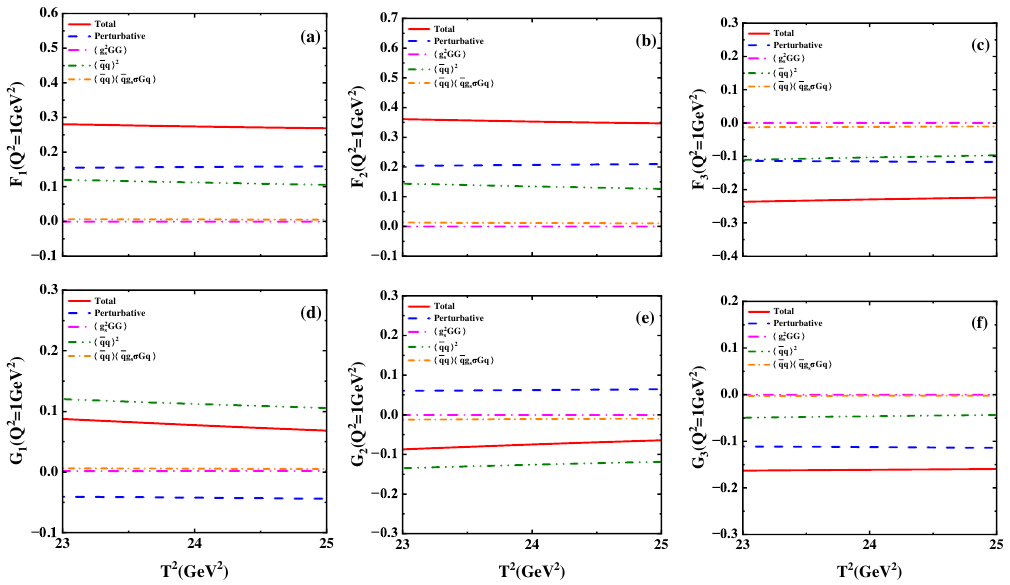}
		\caption{The contributions of the perturbative part and different vacuum condensate terms with variation of Borel parameter $T^2$ for the $\Sigma_b\to\Sigma_c$ transition form factors.}
		\label{BWS}
\end{figure*}
	
\begin{figure*}
		\centering
		\includegraphics[width=18cm]{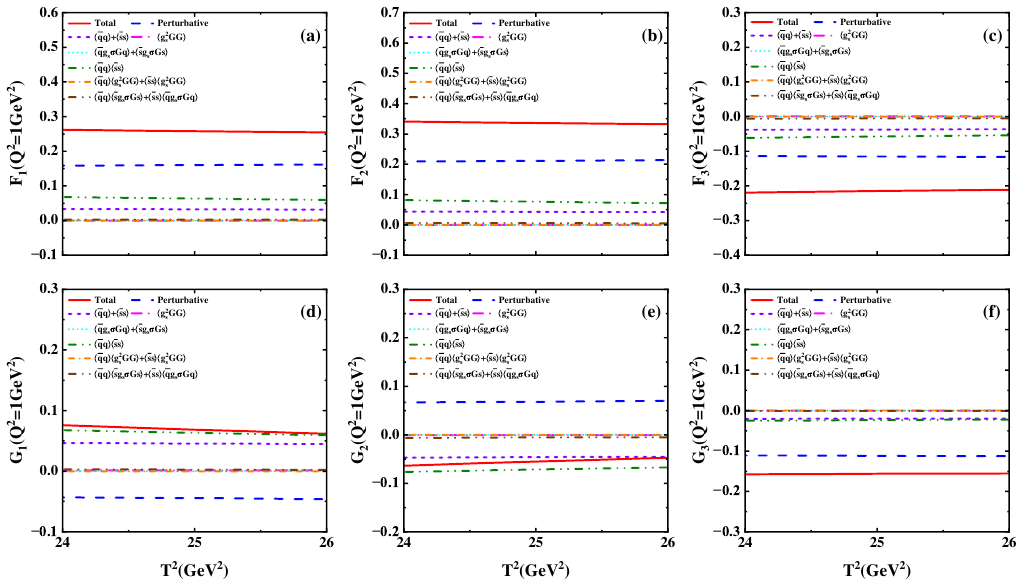}
		\caption{It is the same as the Fig.~\ref{BWS}, but for the $\Xi'_b\to\Xi'_c$ transition form factors.}
		\label{BWXp}
\end{figure*}

\begin{figure*}
	\centering
	\includegraphics[width=18cm]{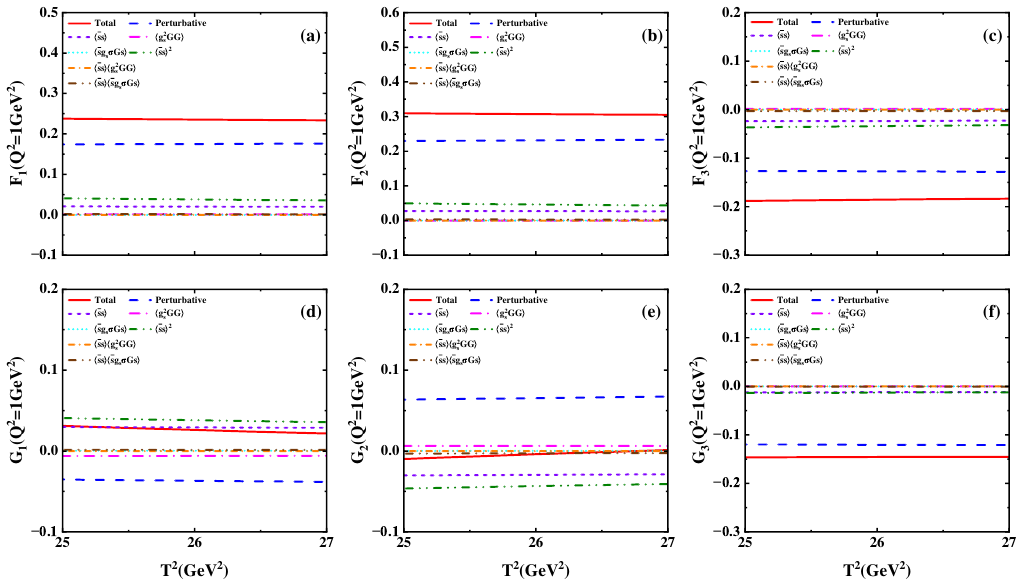}
	\caption{It is the same as the Fig.~\ref{BWS}, but for the $\Omega_b\to\Omega_c$ transition form factors.}
	\label{BWO}
\end{figure*}

\section{The fitting parameters for the upper and lower bounds of the form factors in $z$ series expand approach.}\label{Sec:AppB}
\begin{table}[htbp]
	\begin{ruledtabular}\caption{Fitting parameters for the upper bounds of form factors (FF) using the $z$-series expansion, and the form factor values at $q^2= q_{\max}^2=(m_{\mathcal{B}_i}-m_{\mathcal{B}_f})^2$.}
		\label{ZPU}
		\renewcommand\arraystretch{1.3}
		\begin{tabular}{c c c c c c}
			Mode&FF&$F(0)$&$a$&$b$&$F(q^2_{\max})$\\ \hline
			\multirow{6}*{$\Sigma_b \to \Sigma_c$}&$F_1$ &$0.32$&$-10.39$&$46.89$&$0.61$ \\
			~&$F_2$ &$0.41$&$-14.27$&$73.62$&$0.85$ \\
			~&$F_3$ &$-0.21$&$-8.56$&$97.66$&$-0.35$\\
			~&$G_1$&$0.080$&$34.65$&$-117.81$&$-0.034$ \\
			~&$G_2$&$-0.056$&$103.01$&$512.30$&$0.19$\\
			~&$G_3$&$-0.14$&$-15.01$&$135.87$&$-0.26$\\ \hline
			\multirow{6}*{$\Xi'_b \to \Xi'_c$}&$F_1$ &$0.31$&$-16.61$&$91.45$&$0.66$ \\
			~&$F_2$ &$0.40$&$-19.17$&$109.37$&$0.90$ \\
			~&$F_3$ &$-0.20$&$-9.49$&$93.52$&$-0.34$ \\
			~&$G_1$&$0.068$&$73.84$&$-477.26$&$-0.11$\\
			~&$G_2$&$-0.039$&$140.17$&$-739.33$&$0.18$\\
			~&$G_3$&$-0.14$&$-12.54$&$102.00$&$-0.25$\\ \hline
			\multirow{6}*{$\Omega_b \to \Omega_c$}&$F_1$ &$0.29$&$-16.08$&$72.68$&$0.62$ \\
			~&$F_2$ &$0.37$&$-15.02$&$59.27$&$0.78$ \\
			~&$F_3$ &$-0.17$&$-2.65$&$6.70$&$-0.26$ \\
			~&$G_1$&$0.030$&$108.71$&$-597.36$&$-0.10$\\
			~&$G_2$&$0.013$&$-4.32\times 10^2$&$2.35\times10^3$&$0.25$\\
			~&$G_3$&$-0.13$&$-7.48$&$27.24$&$-0.22$
		\end{tabular}
	\end{ruledtabular}
\end{table}

\begin{table}[htbp]
	\begin{ruledtabular}\caption{Fitting parameters for the lower bounds of form factors (FF) using the $z$-series expansion, and the form factor values at $q^2= q_{\max}^2=(m_{\mathcal{B}_i}-m_{\mathcal{B}_f})^2$.}
		\label{ZPL}
		\renewcommand\arraystretch{1.3}
		\begin{tabular}{c c c c c c}
			Mode&FF&$F(0)$&$a$&$b$&$F(q^2_{\max})$\\ \hline
			\multirow{6}*{$\Sigma_b \to \Sigma_c$}&$F_1$ &$0.25$&$-10.41$&$60.52$&$0.47$ \\
			~&$F_2$ &$0.33$&$-13.95$&$82.62$&$0.67$ \\
			~&$F_3$ &$-0.26$&$-5.76$&$54.65$&$-0.41$\\
			~&$G_1$&$0.061$&$90.57$&$-483.83$&$-0.17$ \\
			~&$G_2$&$-0.078$&$40.53$&$-126.55$&$0.058$\\
			~&$G_3$&$-0.19$&$-10.22$&$72.25$&$-0.32$\\ \hline
			\multirow{6}*{$\Xi'_b \to \Xi'_c$}&$F_1$ &$0.24$&$-13.49$&$74.60$&$0.48$ \\
			~&$F_2$ &$0.31$&$-16.51$&$94.40$&$0.66$ \\
			~&$F_3$ &$-0.25$&$-13.27$&$111.52$&$-0.47$ \\
			~&$G_1$&$0.055$&$115.25$&$-741.08$&$-0.18$\\
			~&$G_2$&$-0.056$&$80.69$&$-436.98$&$0.12$\\
			~&$G_3$&$-0.19$&$-19.68$&$150.16$&$-0.39$\\ \hline
			\multirow{6}*{$\Omega_b \to \Omega_c$}&$F_1$ &$0.22$&$-13.38$&$54.22$&$0.45$ \\
			~&$F_2$ &$0.29$&$-13.43$&$47.26$&$0.60$ \\
			~&$F_3$ &$-0.22$&$-6.77$&$35.48$&$-0.37$ \\
			~&$G_1$&$0.015$&$196.84$&$-699.31$&$-0.12$\\
			~&$G_2$&$3.18\times10^{-9}$&$-1.25\times10^{9}$&$4.55$&$0.19$\\
			~&$G_3$&$-0.18$&$-13.81$&$78.22$&$-0.34$
		\end{tabular}
	\end{ruledtabular}
\end{table}

\bibliography{ref.bib}
\end{document}